\documentclass[aps,prb,twocolumn,amsfonts,superscriptaddress,showpacs,final,floatfix,preprintnumbers]{revtex4-1}

\usepackage{amsmath, amssymb, mathrsfs}
\usepackage{latexsym}
\usepackage{graphicx,epstopdf,color}
\usepackage{braket}
\usepackage{subfigure}
\usepackage{times}
\usepackage{bm}

\usepackage[breaklinks]{hyperref}
\hypersetup{
   colorlinks,
   citecolor=blue,
   filecolor=blue,
   linkcolor=blue,
   urlcolor=blue
}

\allowdisplaybreaks[1]

\begin{document}

\title{Floquet Engineering of Haldane Chern Insulators and Chiral bosonic phase transitions}
\author{Kirill Plekhanov}
\affiliation{LPTMS, CNRS, Univ. Paris-Sud, Universit\'e Paris-Saclay, 91405 Orsay, France}
\affiliation{Centre de Physique Th\'eorique, Ecole Polytechnique, CNRS, Universit\'e Paris-Saclay, F-91128 Palaiseau, France}
\author{Guillaume Roux}
\affiliation{LPTMS, CNRS, Univ. Paris-Sud, Universit\'e Paris-Saclay, 91405 Orsay, France}
\author{Karyn Le Hur}
\affiliation{Centre de Physique Th\'eorique, Ecole Polytechnique, CNRS, Universit\'e Paris-Saclay, F-91128 Palaiseau, France}
\date{\today}

\begin{abstract}
The realization of synthetic gauge fields has attracted a lot of attention recently in relation with periodically driven systems and the Floquet theory. In ultra-cold atom systems in optical lattices and photonic networks, this allows to simulate exotic phases of matter such as quantum Hall phases, anomalous quantum Hall phases and analogs of topological insulators. In this paper, we apply the Floquet theory to engineer anisotropic Haldane models on the honeycomb lattice and two-leg ladder systems. We show that these anisotropic Haldane models  still possess a topologically non-trivial band structure associated with chiral edge modes (without the presence of a net unit flux in a unit cell), then referring to the quantum anomalous Hall effect. Focusing on (interacting) boson systems in s-wave bands of the lattice, we show how to engineer through the Floquet theory, a quantum phase transition between a uniform superfluid and a BEC (Bose-Einstein Condensate) analog of FFLO (Fulde-Ferrell-Larkin-Ovchinnikov) states, where bosons condense at non-zero wave-vectors. We perform a Ginzburg-Landau analysis of the quantum phase transition on the graphene lattice,  and compute observables such as chiral currents and the momentum distribution. The results are supported by exact diagonalization calculations and compared with those of the isotropic situation. The validity of high-frequency expansion in the Floquet theory is also tested using time-dependent simulations for various parameters of the model. Last, we show that the anisotropic choice for the effective vector potential allows a bosonization approach in equivalent ladder (strip) geometries.
\end{abstract}

\maketitle

\section{Introduction}

Topological phases of matter play an important role, from quantum Hall physics \cite{Klitzing1980qhe, Tsui1982fqhe, Laughlin1981qhe} to the concept of symmetry protected topological phases \cite{Haldane1983symProtTopPhases, Affleck1987symProtTopPhases, Affleck1989symProtTopPhases, Chen2013sptCohomology, Senthil2015symProtTopPhasesReview}. Topological Bloch bands have also been detected in quantum materials (topological insulators) \cite{KaneMele2005model,Konig2007HgTeExp, HasanKane2010TopInsulators, Qi2011topInsReview, Fruchart2013topInsReview, Bernevig2013Book}, ultra-cold atoms \cite{Dalibard2011topoColdAtomsReview, Goldman2014topoColdAtomsReview, Goldman2016topoColdAtomsReview} and photon systems \cite{Carusotto2013topoPhotonics, Lu2014topoPhotonics, LeHur2016topoPhotonics, Hartmann2016qSimIntPhotons}.

Very recently, experimental realizations of the Harper-Hofstadter model
\cite{Harper1955, Hofstadter1976} were performed with ultracold atoms by using the laser-assisted tunneling technique \cite{Aidelsburger2011ArtificialMagField, Aidelsburger2013hofstadterCldAtms, Miyake2013hofstadterCldAtms, Aidelsburger2015hofstadterCldAtms}, proposed for the first time in Ref.~\onlinecite{JakschZoller2003LaserAssistedTunelling}. At the same time, the possibility to generate artificial gauge fields was transferred into the domain of electromagnetic waves, giving opportunity to realize even more exotic photonic analogues of quantum Hall effect \cite{Haldane2008photonicQheTheo, Raghu2008photonicQheTheo}. Few years later, such objects were realized experimentally using the lattice of ferrite rods \cite{Wang2009photonicQheExp} and in a system of coupled optical-ring resonators \cite{Hafezi2013photonicQheHofstExp}.
In a similar way, the artificial magnetic fields realisation of the effective Haldane model \cite{Haldane1988model}, known as the model for the quantum anomalous Hall effect, was done first by using the array of evanescently coupled helical waveguides with the propagation coordinate $z$ playing the role of time \cite{Rechtsman2013FloquetHelicalWaweGuide}, and later in the system of shaken honeycomb optical lattice \cite{Jotzu2014haldaneShaking}, following the first experimental realization of artificial graphene with cold atoms \cite{SoltanPanahi2011artificialGraphene, Tarruell2012honeycombShaking}. This occurred in parallel to first observations of the quantum anomalous Hall effect in quantum materials \cite{Chang2013anQheQMat} (an overview on recent progress in this domain can be found in Refs.~\onlinecite{Wang2015anQheQMatRev, Liu2015anQheQMatRev}). Another experiment with ultracold atoms in triangular flux lattice also succeeded in realizing Ising-XY spin-models \cite{Struck2013isingXyTriangShaking}. Recent experimental realization of Floquet engineered bands in tunable honeycomb lattices with full momentum-resolved measurement of the Berry curvature was also reported in Ref.~\onlinecite{Flaschner2016coldAtomsHaldaneMeasureBerryCurv}.

Importantly, experimental realizations of artificial magnetic fields in cold atom systems and in photonic lattices are based on the application of a periodic time-dependent perturbation. In the case of a non-resonant coupling, the Floquet theory is often used to solve the problem. According to the Floquet theory \cite{Shirley1965Floquet, Sambe1973Floquet, Gesztesy1981Floquet}, one expects that the dynamics of such systems in the high-frequency regime will be separated into the "slow" and "fast" parts. The slow dynamics will thus be simply described in terms of a time-independent effective (Floquet) Hamiltonian. Generally, an exact expression of the effective (Floquet) Hamiltonian is not accessible and one should rely on approximations such as, for example, the Magnus expansion \cite{Magnus1954Expansion, Blanes2009MagnusExpansion} or the high frequency expansion \cite{Maricq1982hfeFloquetNmrOfSolids, Grozdanov1988hfeFloquet, Rahav2003HighFreqExp1, Rahav2003HighFreqExp2} (see Refs.~\onlinecite{Dalibard2014PeriodDrivenQSystems, Eckardt2015FloquetTh, Bukov2015Review} for general review and comparison of the two approaches). These expansions are perturbative and their convergence is not ensured even in the relatively simple case of non-interacting systems. When interactions must be taken into account, the situation becomes even more complicated. Behaviour of driven many-body systems was studied theoretically and numerically \cite{Prosen1998FloquetManyBody, Prosen1998FloquetManyBody2, Prosen1999FloquetManyBody3, Prosen2011floquetXYchain, Alessio2013ManyBodyPdsEnLocalization, Alessio2014LongTBehavPDInterSyst, Lazarides2014PeriodThermod, Lazarides2015floquetManyBodyLclzd, Ponte2015FloquetManyBodyLocaliz, Abanin2015periodManyBodySysts, Abanin2015floquetManyBody, Genske2015FloquetBoltzmannEquation, Mori2016periodManyBodySystems, Kuwahara2016floquetMagnusHeating} and the invalidity of the high frequency expansions was often related to the problem of heating in many-body systems \cite{Eckardt2008DressedMatterWaves, Heyl2010PerDrivKondo, Russomanno2012FloqHeatInteg, Alessio2013ManyBodyPdsEnLocalization, Alessio2014LongTBehavPDInterSyst, Lazarides2014PeriodThermod, Lazarides2015floquetManyBodyLclzd, Abanin2015periodManyBodySysts, Abanin2015floquetManyBody, Citro2015ManyBodyKapitza, Ponte2015ManyBodyLocalizInPDS, Bilitewski2015floquetHeating, Genske2015FloquetBoltzmannEquation, Bukov2015HeatingPdsTwoBands, Mori2016periodManyBodySystems, Kuwahara2016floquetMagnusHeating}.

Nevertheless, today a lot of interest is dedicated to "Floquet engineering"~\cite{Dalibard2014PeriodDrivenQSystems, Bukov2015Review, Goldman2015PeriodDrivenQSystems, Eckardt2015FloquetTh}. More generally, it consists in generating an effective (Floquet) Hamiltonian with desired properties, starting from a more trivial one by superimposing a periodic time-dependent perturbation. Using the Floquet theory (and the high-frequency expansion) and a well chosen geometry allows to design targetted Hamiltonians.
Many theoretical studies were done in order to investigate the possibility to obtain Floquet topological (Chern) insulators from irradiated graphene structures \cite{Oka2009FloquetRadiation, Calvo2011FloquetRadiation, Kitagawa2011photoinducedQheTransport, Morell2012floquetirridiatedGraphene, Lindner2011quantumWellsFloquet, Lindner2013quantumWellsFloquet, Cayssol2013FloquetTopIns, Delplace2013FloquetRadiation, Grushin2014FloquetRadiation, Gomez2014FloquetRadiation}, that were recently observed experimentally \cite{Wang2013FloquetRadiation}. 
Driven superconducting systems were considered in order to observe 
Floquet Majorana fermions \cite{Jiang2011FloquetMajorana, Liu2012FloquetMajorana, Tong2013FloquetMajorana, Thakurathi2013FloquetMajorana, Kundu2013floquetMajorana, Wang2015FloquetMajorana}.
These works are in deep connection with the study of topological properties and corresponding topological invariants of periodically driven systems
\cite{Kitagawa2010FloquetTopInvar, Rudner2013FloquetBulkEdgeCorresp, Gomez2013FloquetTopInvar, Lababidi2014FloquetTopInvar, Perez2014FloquetTopInvar, Carpentier2015FloquetTopIndex, NathanRudner2015FloqTopInsClass, Potter2016floqTopPreiodicTable1D, Maczewsky2016FloqAnomTopInsExp}, showing that their topological structure is even richer then in the general classification at equilibrium \cite{Kitaev2009periodicTable}.

Based on recent theoretical proposals \cite{Fang2012photonicEffMagField, Fang2013photonicHaasVanAlphen} and experimental realizations \cite{Roushan2016supercondTriang}, in this work we investigate the possibility to simulate quantum anomalous Hall effect for photons. We consider the model for bosons defined on the honeycomb lattice, with the Hamiltonian comprising the term of periodically modulated nearest neighbor hoppings. In the high frequency regime, the time-dependent term does not contribute in average, but creates effective hoppings between next nearest neighbours with purely imaginary amplitudes, leading to the generation of an anisotropic version of the Haldane model. We check the validity of the high frequency expansion by performing numerical comparison of the time-evolution with exact and effective Hamiltonians and we infer the regime of consistency between the two Hamiltonians.

In the case of cold atoms systems (or also exotic polariton superfluids \cite{Amo2009polaritonSuperfluid} and photonic BECs \cite{Klaers2010photonBec}), the generation of the Haldane model, reported in Refs.~\onlinecite{Jotzu2014haldaneShaking, Flaschner2016coldAtomsHaldaneMeasureBerryCurv}, allows one to investigate the rich properties of lattice bosons subjected to artificial gauge fields as performed in 1D \cite{Struck2012artGaugeFields1Dfmbec, Jimenez-Garcia2012artGaugeFields1D} and ladder \cite{Atala2014bosonicLaddersUltracoldAtoms} systems and on triangular lattices \cite{Struck2011frustratedMagnetismeTriangLattices, Struck2013isingXyTriangShaking}. It is known that this can give rise to the condensation phenomenon of bosons at non-zero momentum \cite{Lim2010fmbec, Moller2010fmbec, VassicPetrescu2015}. This effect is similar to the FFLO phase \cite{FuldeFerell1964fflo_ff, FuldeFerell1964fflo_lo}, and can be realized by working only with s-wave bands of the optical lattice. Recently, square lattice realization of this effect was performed on a 2D lattice \cite{Kennedy2015fmBecHofstadterFloquet}. Finite momentum BEC was predicted \cite{Isacsson2005fmBecPiOrbsTheo,Kuklov2006fmBecPiOrbsTheo,Liu2006fmBecPiOrbsTheo} and observed with p-wave band superfluids \cite{Muller2007fmBecPiOrbsExp, Wirth2011fmBecPiOrbsExp}. Related experimental works were also performed recently, based on a near-resonant hybridization of s- and p-bands of bosons \cite{Parker2013fmbec, Khamehchi2016fmbec}. In this paper, we show that the configuration of artificial magnetic fields, appearing in the effective anisotropic Haldane model for bosons in s-wave bands, also supports the finite momentum phase with two-well structure, and that the transition between the zero momentum and the finite momentum phase is of second order, in contrast to the isotropic situation where chiral currents show a discontinuity at the phase transition \cite{VassicPetrescu2015}. We show moreover that the nature of this transition changes in the ladder analogue of the model, giving rise to the apparition of flat bands. The role of interactions in the ladder geometry far from the flat band region is then studied using bosonization \cite{Haldane1981a, Gogolin2004bosonization, Giamarchi2003QpIn1D, Cazalilla2011OneDimBosons}, in relation with
current experimental developments \cite{Atala2014bosonicLaddersUltracoldAtoms}.

This work is organized in the following way: in Sec.~\ref{sec:artMagFieldsAndEffModels} we describe the initial model with time-dependent Hamiltonian. In Sec.~\ref{sec:FloquetTheory} we recall some details of the Floquet theory and of the related high frequency expansion, used in Sec.~\ref{sec:effHaldaneModel_creation} to construct the effective static Hamiltonian, corresponding to the anisotropic Haldane model for photons. We outline properties of the anisotropic Haldane model related to topological phase transition, that could be accessed with photons, in Sec.~\ref{sec:effHaldaneModel_edgeModes}. In Sec~\ref{sec:enginFmbec} we discuss the groundstate properties of the anisotropic Haldane model for bosons, that could be observed in cold atom systems. We characterize the phase corresponding to the Bose-Einstein condensate at zero momentum (Secs.~\ref{sec:zmGs}, \ref{sec:zmBglbv}) and the phase at finite momentum, as well as their quantum phase transition (Secs.~\ref{sec:sfToCsf_t2crit}, \ref{sec:fmPhase}). Observables such as chiral currents and momentum distribution are studied. The role of moderate interactions on the quantum phase transition is discussed in Sec. \ref{sec:sfToCsf_ints}. Then, we perform the numerical analysis of the high frequency expansion in Sec.~\ref{sec:numFloquetCheck}. 
Finally, in Sec.~\ref{sec:ladders} we study a ladder analogue of the anisotropic Haldane model. In this section we also mention some other ladder models that can be generated with our approach on different types of lattices. Extra technical details will be provided in Appendices. Appendix \ref{app:DysonPerturbationTheory} will discuss the Floquet theory and the Dyson expansion to compute observables. In Appendix \ref{app:fmPhaseInter}, we discuss in more details the Ginzburg-Landau expansion in the FM phase in the presence of interactions.

\section{Effective models with artificial gauge fields in Floquet lattices \label{sec:artMagFieldsAndEffModels}}

The system is described by the Hamiltonian $\hat{H}(t) = \hat{H}_0 + \hat{V}(t)$, defined on a general bipartite lattice where two sublattices are denoted by $A$ and $B$. The Planck constant $\hbar$ is set to unity for simplicity.
\begin{align}
\label{eq:hamiltonian_timeDep_1p}
&\hat{H}_0 =
    \omega_A \sum\limits_{i\in A} \hat{a}^\dag_i \hat{a}_i + 
    \omega_B \sum\limits_{j\in B} \hat{b}^\dag_j \hat{b}_j - 
    t_1 \sum\limits_{\braket{ij}} \left(
    \hat{a}^\dag_i \hat{b}_j + \hat{b}^\dag_j \hat{a}_i 
    \right)
\notag \\
&\hat{V}(t) =
	V \sum\limits_{\braket{ij}}
    \cos(\omega t + \theta_{ij}) 
    \left( \hat{a}^\dag_i \hat{b}_j + \hat{b}^\dag_j \hat{a}_i
    \right)\;.
\end{align}
Here $\hat{a}_i(\hat{b}_j)$ are annihilation operators on site $i(j)$ of sublattice $A(B)$, and $\omega_A (\omega_B)$ is the corresponding frequency in the case of a photonic optical lattice.

Sublattices are coupled between them through a constant tunnelling term $t_1$ and a time-dependent term $V$ (which is periodic in time) coupling nearest neighbours. In cavity systems this consists to couple the cavities with a DC and AC terms and assuming that each of them has approximately the same length. $\theta_{ij} = - \theta_{ji}$ are scalar phases associated to each oriented link between sites $i$ and $j$, and $\sum_{\Braket{ij}}$ denotes the summation over all possible pairs of nearest neighbours (NN) with $i$ sitting on the sublattice $A$ and $j$ -- on the sublattice $B$. Related scenarios have been suggested to realize quantum Hall phases of bosons \cite{Hayward2012fQheJaynesCummingsHubbard}.

The laboratory frame realization of such Hamiltonian was proposed in the context of photonic lattices \cite{Fang2013photonicHaasVanAlphen, Fang2012photonicEffMagField}. It was also successfully implemented using superconducting qubits arranged in a triangular loop with pairwise couplings \citep{Roushan2016supercondTriang}, based on earlier theoretical suggestions \cite{Koch2010circuitQedPhotonLattices}. In the following sections we will show that this approach can be used to generate the model for quantum anomalous Hall effect.

Similar systems were also experimentally realized with ultracold atoms, by using the lattice shaking \cite{Jotzu2014haldaneShaking} or laser assisted tunneling techniques \cite{Aidelsburger2013hofstadterCldAtms, Miyake2013hofstadterCldAtms, Aidelsburger2015hofstadterCldAtms}. Related theoretical proposal was considered in Ref.~\onlinecite{Zheng2014shakingHaldaneTheo}. Photon-assisted tunneling was also suggested to generate artificial magnetic fields in Ref.~\onlinecite{Kolovsky2011photonAssistedTunelling}. In these cases we should however think of $\omega_A$ and $\omega_B$ as of chemical potentials on two sublattices. It was argued that after moving to the rotating frame (and applying the Floquet theory in the non-resonant case), one can eventually generate artificial magnetic fields. In Ref.~\onlinecite{Jotzu2014haldaneShaking} this was used to effectively generate the anisotropic Haldane model. Within these protocols, we will show that we can engineer a BEC analogue of FFLO states, where bosons (in the s-wave band of the optical lattice) condense at a non-zero wave vector.

Inspired by previous works~\cite{VassicPetrescu2015, Lim2010fmbec}, we are also interested in rich properties of the many-body system in the interacting regime. We consider the simplest possible case of interactions, described by the Hamiltonian term
\begin{equation}
\hat{H}_\text{BH} = 
\frac{U}{2} \sum\limits_{i \in A \oplus B} \hat{n}_i \left( \hat{n}_i - 1 \right)
\end{equation}
where $U$ is the strength of repulsive on-site Bose-Hubbard interactions and $\hat{n}_i$ is the number operator on site $i$.
In the system of single-component bosons such interactions can originate from the \textit{s}-wave collisions between atoms. An experimental realization of the Bose-Hubbard model can also be achieved in photonic lattices in the highly detuned limit, with an (approximately exact) mapping between the Jaynes-Cummings lattice model and the Bose-Hubbard model. This problem was addressed in particular in Refs.~\onlinecite{Boissonneault2009dispersiveRegimeCircuitQed, Hoffman2011photonPhotonInteractions}.

The effect of interactions in the periodically driven systems is highly non-trivial. It was argued that it can lead to the problem of heating, indicating the non-unitarity of the time-evolution \cite{Eckardt2008DressedMatterWaves, Heyl2010PerDrivKondo, Russomanno2012FloqHeatInteg, Alessio2013ManyBodyPdsEnLocalization, Alessio2014LongTBehavPDInterSyst, Lazarides2014PeriodThermod, Lazarides2015floquetManyBodyLclzd, Abanin2015periodManyBodySysts, Abanin2015floquetManyBody, Citro2015ManyBodyKapitza, Ponte2015ManyBodyLocalizInPDS, Bilitewski2015floquetHeating, Genske2015FloquetBoltzmannEquation, Bukov2015HeatingPdsTwoBands, Mori2016periodManyBodySystems, Kuwahara2016floquetMagnusHeating}. The main goal of this paper is to considerate weak to moderate interaction effects and understand the emergent quantum phase transition from the uniform BEC to the FFLO-type BEC almost in a single-particle manner. Interaction effects will also be discussed using ED (for a small number of particles) and using a Ginzburg-Landau approach (for an infinite number of particles). In an analogue ladder geometry, the effect of interactions will be captured using the Haldane representation of bosons \cite{Haldane1981a, Gogolin2004bosonization, Giamarchi2003QpIn1D, Cazalilla2011OneDimBosons}, which is
reminiscent of the Gross-Pitaevski representation in the two dimensional system. The main difference will be the nature of the quantum phase transition: in the ladder system, there will be the occurrence of flat bands in the energy spectrum in the vicinity of the quantum phase transition.

\section{Floquet theory. Long time dynamics of periodically driven systems \label{sec:FloquetTheory}}

We consider the system evolving with the general periodic time-dependent Hamiltonian $\hat{H}(t)$ with the modulation frequency $\omega = 2 \pi/T$.
According to the Floquet theory \cite{Shirley1965Floquet, Sambe1973Floquet, Gesztesy1981Floquet}, one can expect that the dynamics of such systems in the high-frequency regime will be separated into the "slow" and "fast" parts. From the point of view of states in the Hilbert space, this means that each eigenstate $\Ket{\psi_n(t)}$ of the Hamiltonian $\hat{H}(t)$ can be rewritten as follows: 
\begin{equation}
\Ket{\psi_n(t)} = e^{-i\epsilon_n t} \Ket{\phi_n(t)}
\end{equation}
where $\Ket{\phi_n(t)}$ are Floquet states (analogues of Bloch states in the time-space) verifying $\Ket{\phi_n(t)} = \Ket{\phi_n(t+T)}$. $\epsilon_n$ are called Floquet quasienergies. 

This decomposition can be used to write the expression of the evolution operator:
\begin{align}
\hat{U} \left( t, t_0 \right) =
& \sum\limits_{n}
	e^{-i \epsilon_n \left(t-t_0\right)}
	\Ket{\phi_n(t)} \Bra{\phi_n(t_0)}
\notag \\ 
\underset{t = t_0+mT}{=}
& \sum\limits_{n}
	e^{-i m \epsilon_n T}
	\Ket{\phi_n(t_0)} \Bra{\phi_n(t_0)}
\end{align}
The "fast" intra-periodic sub-motions of the system are described by the term $\Ket{\phi_n(t)} \Bra{\phi_n(t_0)}$, that are often associated with "Kick" operators $\hat{K}_\text{eff} \left( t \right)$ and $\hat{K}_\text{eff} \left( t_0 \right)$ \citep{Dalibard2014PeriodDrivenQSystems}. The "slow" long-time dynamics is due to the term $e^{-i \epsilon_n \left(t-t_0\right)}$, that can also be interpreted as the evolution under some static effective (Floquet) Hamiltonian $\hat{H}_\text{eff}$. This leads to the following rewriting:
\begin{equation}
\hat{U} \left( t, t_0 \right) = 
e^{-i\hat{K}_\text{eff}\left( t \right)}
e^{-i\hat{H}_\text{eff}\left( t-t_0 \right)}
e^{ i\hat{K}_\text{eff}\left( t_0 \right)}
\end{equation}

In general case one can not deduce an exact expression of the effective (Floquet) Hamiltonian and should rely on approximations such as, for example, the Magnus expansion (ME) \cite{Magnus1954Expansion, Blanes2009MagnusExpansion} or the high frequency expansion (HFE) \cite{Maricq1982hfeFloquetNmrOfSolids, Grozdanov1988hfeFloquet, Rahav2003HighFreqExp1, Rahav2003HighFreqExp2} (see Refs.~\onlinecite{Dalibard2014PeriodDrivenQSystems, Eckardt2015FloquetTh, Bukov2015Review} for general reviews and comparison of these two approaches). In particular, up to the first order term, the HFE can be expressed as follows:
\begin{align}
\label{eq:effectiveHamiltonian}
& \hat{H}_\text{eff} =
\hat{H}^{(0)} + \sum\limits_{j = 1}^\infty
	\frac{1}{j\omega} \left[ \hat{H}^{(j)}, \hat{H}^{(-j)} \right] + \dots
\notag \\
& \hat{K}_\text{eff}(t) =
\frac{1}{i \omega} \sum\limits_{j = 1}^\infty
	\left( \frac{e^{ij\omega t}}{j} \right)
	\hat{H}^{(j)} + \dots
\end{align}
where $\hat{H}^{(j)}$ are distinct Fourier components of $\hat{H}(t)$.

\section{Effective anisotropic Haldane model}

\subsection{Generating effective anisotropic Haldane model \label{sec:effHaldaneModel_creation}}

\begin{figure}
\includegraphics[width=0.48\textwidth]
{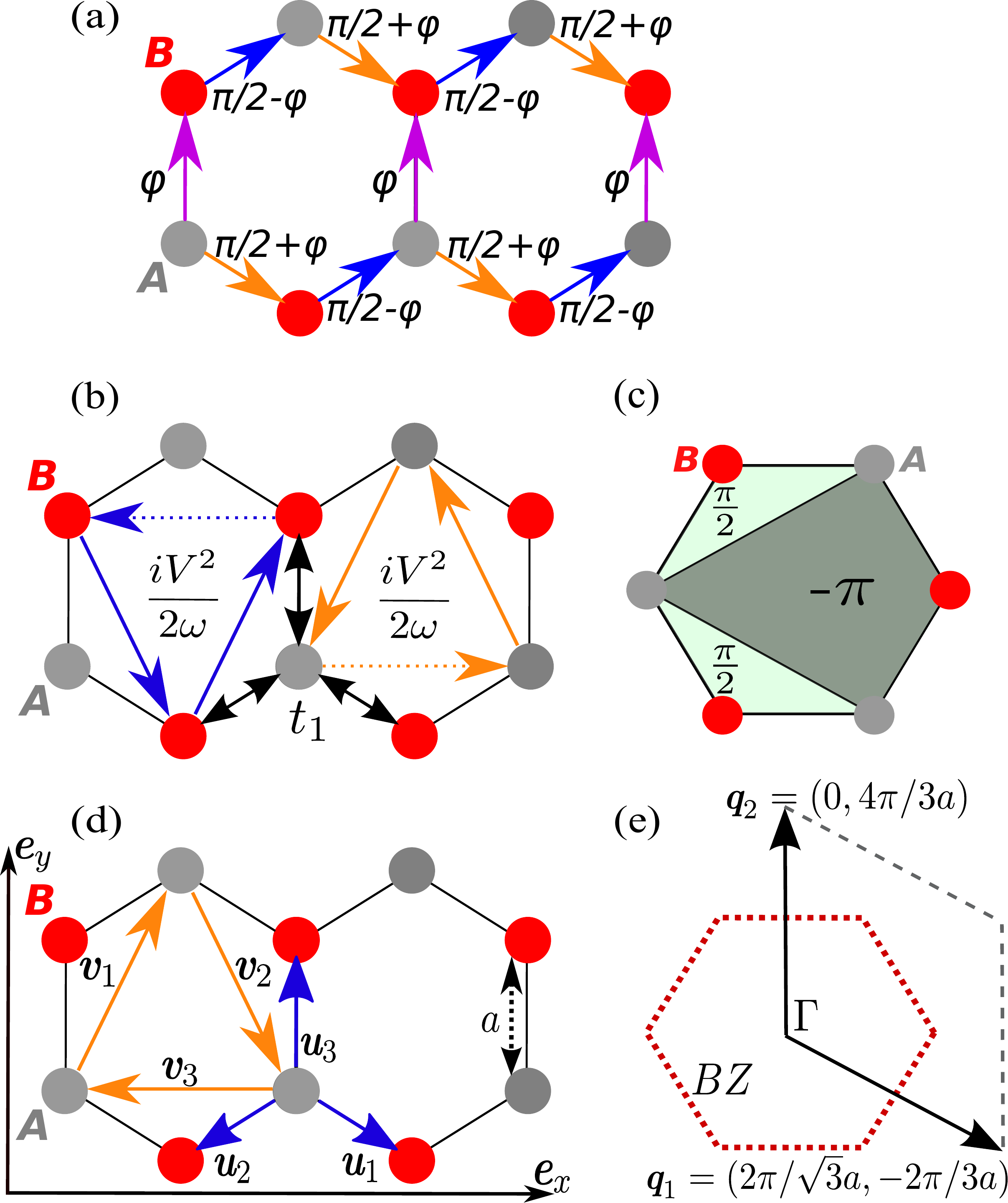}
\caption{\textbf{(a)} Distribution of phases on the NN bonds in the time-dependent Hamiltonian \eqref{eq:hamiltonian_timeDep_1p} required to generate effective anisotropic Haldane model. The phase $\varphi$ is the free parameter. We notice that these phases only change the value of the NNN hopping amplitude and has no effect on the Haldane phase, that is always equal to $\pi/2$. \textbf{(b)} Anisotropic Haldane model. Coloured bold lines represent NNN hoppings with amplitude modulus $\frac{V^2}{2 \omega}$ and assigning a phase $\pm\pi/2$ if they are made in the anticlockwise (clockwise) direction. Dotted lines correspond to the hopping absent in the model. \textbf{(c)} Magnetic flux distribution in the anisotropic Haldane model. The total magnetic flux through the unit cell sums up to zero, with closed loops of vector potentials involving kite geometries rather then triangles as in the isotropic case. \textbf{(d)} Definition of vectors $\bm{u}_i$ and $\bm{v}_i$ on the honeycomb lattice. Each hexagon has the length of the side equal to $a$. \textbf{(e)} Reciprocal lattice vectors $\bm{q}_1$, $\bm{q}_2$ and the first Brillouin zone (BZ).}
\label{fig:anisHaldane_schemes}
\end{figure}

We now apply the Floquet formalism described in the previous subsection to the Hamiltonian of Eq.~\eqref{eq:hamiltonian_timeDep_1p}. In order to describe the long-time dynamics of the system, we use Eq.~\eqref{eq:effectiveHamiltonian} to derive the effective Hamiltonian $\hat{H}_\text{eff}$. In the following, we will also denote by 
$\hat{H}^{(i)}_\text{eff}$ contributions to $\hat{H}_\text{eff}$ at the order 
$1/\omega$ in such a way that $\hat{H}_\text{eff} = \sum_i \hat{H}^{(i)}_\text{eff}$.
The zero order contribution reads: $\hat{H}^{(0)}_\text{eff} = \hat{H}_0$.

At the order one in the HFE we obtain:
\begin{align}
\hat{H}^{(1)}_\text{eff} = &
\frac{i V^2}{2 \omega} 
\sum\limits_{\Braket{\Braket{ik}}}
	\sin \left( \Theta_{ik} \right)
	\left( 
		\hat{a}^\dag_i \hat{a}_k - \hat{a}^\dag_k \hat{a}_i 
	\right)
\notag \\
& -\frac{i V^2}{2 \omega} 
\sum\limits_{\Braket{\Braket{jl}}}
	\sin \left( \Theta_{jl} \right)
	\left( 
		\hat{b}^\dag_j \hat{b}_l - \hat{b}^\dag_l \hat{b}_j 
	\right) \;.
\end{align}
We have defined $\Theta_{ik} = \theta_{ij} + \theta_{jk}$ for each couple of next (second) nearest neighbours (NNN) $i$ and $k$ sharing both the same NN with the index $j$. We have denoted by $\sum\limits_{\Braket{\Braket{ik}}}$ the summation over all NNN on either sublattice $A$ or sublattice $B$.

We see that terms in $\hat{H}^{(1)}_\text{eff}$ are purely imaginary NNN hoppings whose amplitude $t_{2,ik}$ depends on the phases $\theta_{ij}$ and $\theta_{jk}$ on corresponding links: 
\begin{equation}
t_{2,ik} = -\frac{V^2}{2 \omega} \sin \left( \Theta_{ik} \right)
\end{equation}
The physical interpretation of this effect can be seen as follows: when the hopping mediated by the time-driving term is performed, the particle acquires an energy $\omega$. Since this energy is huge, the corresponding state is highly unstable and the particle is forced to perform the conjugated NN hopping and re-emits the energy $\omega$. Since these hoppings are dephased in our model, this leads to the apparition of the effective phase acquired by the particle, resulting in the generation of an artificial gauge field.

In order to be able to solve the problem, one should consider a particular choice of the lattice geometry and of the phases $\theta_{ij}$. By having in mind the idea of obtaining a topologically non-trivial model with non zero Chern number, we decided to explore the possibility of generating an effective Haldane model \cite{Haldane1988model}.
We thus consider the problem defined on the honeycomb lattice from now on. Fig.~\ref{fig:anisHaldane_schemes}(a) represents a possible choice of phases $\theta_{ij}$ that fulfills our requirement. The resulting term in the effective Hamiltonian corresponds to the anisotropic Haldane model with absent horizontal NNN hoppings and with amplitude  $t_2 = -V^2/2\omega$ of four leftover NNN hoppings at each unit cell:
\begin{equation}
\hat{H}^{(1)}_\text{eff} =
-t_2 \Big(
\sum\limits^\text{anis}_{\Braket{\Braket{ik}}}
	e^{\pm i\pi/2}\hat{a}^\dag_i \hat{a}_k + 
\sum\limits^\text{anis}_{\Braket{\Braket{jl}}}
	e^{\pm i\pi/2}\hat{b}^\dag_j \hat{b}_l + \text{h.c.}
\Big)
\end{equation}
In the last equation, we denoted the presence of the anisotropy by the label \textit{anis} above the summation sign.
The configuration of hopping amplitudes in the anisotropic model is also displayed on Fig.~\ref{fig:anisHaldane_schemes}(b). The total magnetic flux through the unit cell of the lattice sums up to zero, such that Landau levels do not appear in the problem. We outline that phases $\theta_{ij}$ modify only the value of NNN hopping amplitudes and has no effect on the Haldane phase, that is always equal to $\pi/2$.

For the purpose of clarifying connections of the HFE to the standard perturbation theory, we perform calculations of some relevant observables (i.e. currents) by using Dyson series \cite{Dyson1949DysonSeries}. We deduce that results of both theoretical approaches coincide in the regime of weak time-dependent perturbation. Details of the calculations can be found in the Appendix \ref{app:DysonPerturbationTheory}. The validity of considering only the first order term in the HFE will also be discussed in Sec.~\ref{sec:numFloquetCheck}.

\subsection{Topological phase transition and chiral edge states \label{sec:effHaldaneModel_edgeModes}}

The Haldane model was first introduced as the model for the quantum anomalous Hall effect for non-interacting fermions at half-filling. In this section we want to show that the same characterization applies to the anisotropic Haldane model, that belongs to the class of Chern insulators and supports topologically protected edge modes. We emphasize that even very weak imaginary NNN hopping amplitude ($t_2 \ll t_1$) will be sufficient to observe edge states. 

First, we fix the notation related to the structure of the Bravais lattice and reciprocal lattice. We define by $\bm{u}_i$, $i\in [1,2,3]$ vectors connecting each site on the sublattice $A$ to its three first neighbours on the sublattice $B$. Then,
$ \bm{v}_k = \frac{1}{2} \sum\limits_{i,j} \xi_{ijk} 
\left( \bm{u}_i - \bm{u}_j \right) $ 
(with $\xi_{ijk}$ -- the Levi-Civita symbol) are vectors connecting NNN sites on the same sublattices, as shown in Fig.~\ref{fig:anisHaldane_schemes}(c). We also consider that the length of the side of each hexagon equals to $a$. 
We notice that any two vectors $\bm{v}_k$ form the basis of the Bravais lattice, that is given by one of sublattices $A$ or $B$.
  
\begin{figure}
\includegraphics[width=0.48\textwidth]
{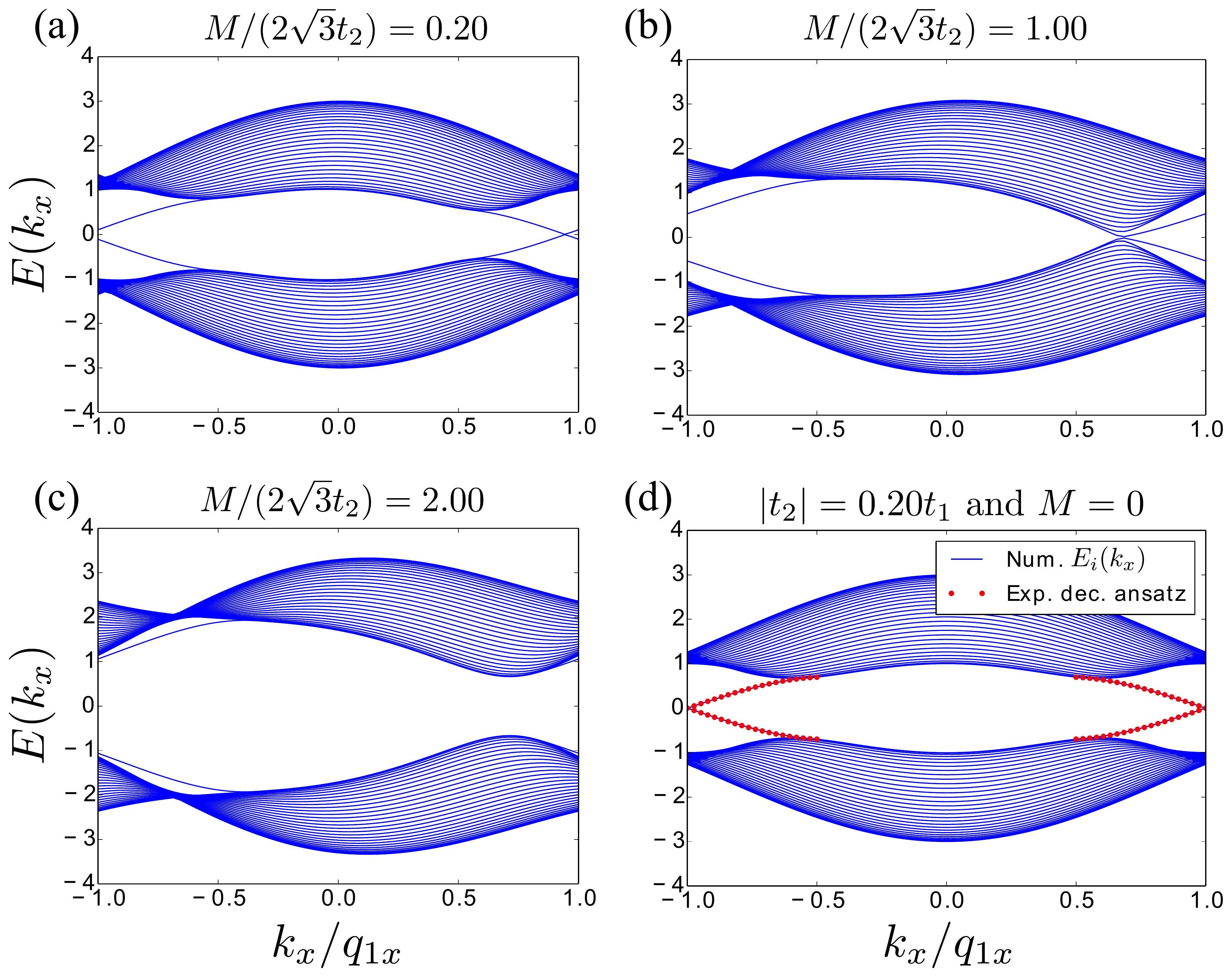}
\caption{\textbf{(a-c)} Numerical calculations of the band structure for the anisotropic Haldane model in the strip geometry with ZigZag edges (cut along the $x$ axis of Fig.~\ref{fig:anisHaldane_schemes}(c)), showing the topological phase transition. Edge modes cross the gap between two bands for $M/t_2 < 2\sqrt{3}$. The gap closes at $M/t_2 = 2\sqrt{3}$. Edge modes do not cross the gap anymore for $M/t_2 > 2\sqrt{3}$. \textbf{(d)} Comparison between numerical results (solid blue lines) and analytical solution for edge modes based on an exponentially decaying ansatz \cite{Petrescu2013kagome, Tianhan2013} (red doted lines). Band structure represented in this way is periodic with period $2q_{1x}$, where $\bm{q}_1$ is the reciprocal lattice vector defined in Fig.~\ref{fig:anisHaldane_schemes}(d).}
\label{fig:topPhaseTransition}
\end{figure}

We remark then that the Hamiltonian
$ \hat{H} = \hat{H}_0 + \hat{H}^{(1)}_\text{eff}$
is easily diagonalized by going to momentum space: 
\begin{equation}
\hat{a}(\hat{b})_{\bm{k}} = \frac{1}{\sqrt{N_c}} 
\sum\limits_{i \in A(B)} e^{-i \bm{k}\cdot \bm{r}_i} \hat{a}(\hat{b})_i
\end{equation}
$N_c$ is the number of unitary cells in the lattice and $\bm{k}$ is the momentum in the first Brillouin zone (BZ), which is spanned by vectors $\bm{q}_1 = (2\pi/\sqrt{3}a,-2\pi/3a)$ and $\bm{q}_2 = (0,4\pi/3a)$ as depicted in Fig.~\ref{fig:anisHaldane_schemes}(d). The Hamiltonian is then rewritten as
\begin{align}
\label{eq:AnisHldn-1pSpct}
\hat{H} &= 
\sum\limits_{\bm{k} \in \mathcal{BZ}}
\hat{\psi}^\dag_{\bm{k}} \cdot
\mathcal{H}(\bm{k}) \cdot
\hat{\psi}_{\bm{k}} 
\notag \\ &=
\sum\limits_{\bm{k} \in \mathcal{BZ}} 
\hat{\psi}^\dag_{\bm{k}} \cdot	
\left[
	\epsilon_0 I_{2\times2}	
	- \bm{d}(\bm{k}) \cdot \bm{\sigma}
\right] \cdot
\hat{\psi}_{\bm{k}}
\end{align}
where
$\sigma_{j}$ are Pauli matrices and $\epsilon_0 = \left( \omega_A + \omega_B \right)/2$. We have also defined the spinor
$ \hat{\psi}_{\bm{k}} = \begin{pmatrix}
	\hat{a}_{\bm{k}} \\
	\hat{b}_{\bm{k}}
\end{pmatrix}$
and three functions of the momentum $\bm{k}$:
\begin{align}
\label{eq:dFunctionsHaldane}
d_x(\bm{k}) &= t_1 \sum_{i=1}^3 \cos(\bm{k} \cdot \bm{u}_i),
\notag \\
d_y(\bm{k}) &= t_1 \sum_{i=1}^3 \sin(\bm{k} \cdot \bm{u}_i),
\\
d_z(\bm{k}) &= - M +2 t_2 \sum_{i=1}^2 \sin(\bm{k} \cdot \bm{v}_i) \;,
\end{align}
with $\bm{u}_i$ and $\bm{v}_i$ the vectors between NN and NNN sites respectively, such that 
$ \bm{v}_k = \frac{1}{2} \sum\limits_{i,j} \xi_{ijk} 
\left( \bm{u}_i - \bm{u}_j \right) $ 
with $\xi_{ijk}$ the Levi-Civita symbol, and $ M = \left( \omega_A - \omega_B \right)/2$ is the Semenoff mass \cite{Semenoff1984model}.
The eigenvalues of $\mathcal{H}(\bm{k})$ read
\begin{align}
\nonumber
\epsilon(\bm{k}) &= \epsilon_0 \pm \left| \bm{d}(\bm{k}) \right| \\
&= \epsilon_0 \pm \sqrt{d^2_x(\bm{k}) + d^2_y(\bm{k}) + d^2_z(\bm{k})}
\end{align}
The corresponding band-structure is illustrated on Fig.~\ref{fig:topPhaseTransition} for various $t_2$ and $M$.

In order to study topological properties of this band-structure, we take the expression of the Berry curvature $B(\bm{k})$ of an energy band:
\begin{equation}
B(\bm{k}) = \frac{\hat{\bm{d}}}{2} \cdot \left(
	\frac{\partial \hat{\bm{d}}}{\partial k_x} \times
	\frac{\partial \hat{\bm{d}}}{\partial k_y}
\right) \quad \text{with }
\hat{\bm{d}} = \frac{\bm{d}(\bm{k})}{\left| \bm{d}(\bm{k}) \right|}\;.
\end{equation}
The flux of the Berry curvature through the BZ is called the Chern number:
\begin{equation}
C_n = \frac{1}{2 \pi} \int\limits_{{BZ}} d^2\bm{k} \ B(\bm{k})
\end{equation}
This quantity expresses how the eigenfunction spinor wraps around the boundary of the BZ. Similarly to the isotropic Haldane model, 
a continuous gauge cannot be defined over the whole BZ, meaning that spinors possess singularities.
It occurs at two Dirac points $\bm{K}_1$ and $\bm{K}_2$. Encircling only these points gives the possibility to estimate the Chern number. We find that the contribution at each point is equal to $\pm 1/2$, where the sign corresponds to the sign of $d_z(\bm{k})$.

The missing link does not affect the possibility of topological bands in the anisotropic Haldane model (thanks to the presence of closed loops of vector potentials involving kite geometries), but modifies the critical point of the topological versus non-topological phase transition and the edge-mode dispersion relation.
More precisely, in the anisotropic Haldane model we find that
$ d_z(\bm{K}_1) = -M - 2 \sqrt{3} t_2 $ and
$ d_z(\bm{K}_2) = -M + 2 \sqrt{3} t_2 $.
Thus, in the regime $t_2 > M/(2\sqrt{3})$, the phase acquired by the particle wrapping around the BZ does not vanish.
In this case, according to the bulk-boundary correspondence \cite{HasanKane2010TopInsulators}, by fixing the chemical potential of the system to the gap between two bands, one should be able to observe edge modes that goes across the boundary of our system in real space. These modes should disappear in the regime $ t_2 < M/(2\sqrt{3})$, after moving through the topological phase transition at $ t_2 = M/(2\sqrt{3})$, for which the gap between two bands closes and the system becomes conducting.

These calculations are verified numerically by diagonalizing the Hamiltonian in the strip geometry, i.e. with periodic boundaries along $x$ and open boundaries along $y$ (ZigZag boundary conditions). The band structure is calculated for different values of $t_2 / M$ and edge modes are detected, as seen on Fig.~\ref{fig:topPhaseTransition}. Moreover, we adapt methods previously used in works of Refs.~\onlinecite{Petrescu2013kagome, Tianhan2013} to derive an analytical calculation of the dispersion relation for edge modes of the anisotropic Haldane model, by considering an exponentially decaying ansatz. In Fig.~\ref{fig:topPhaseTransition}(d), we compare the analytical and numerical solutions.

We outline that if $t_2 > M/(2\sqrt{3})$, even weak values of $t_2$ (compared to $t_1$) and, as a consequence, small amplitudes of the Floquet perturbation $V$, permit to observe edge modes. However, as we will show further in this work, some other interesting properties of the system (not related to the topology in the sense considered in the current section) are observed if we consider the regime when $t_2$ becomes of the order of $t_1$ and when interactions and many-body effects are also taken into account.
This means, in particular, that in the limit $\omega \rightarrow \infty$, relevant for the validity of the HFE, $V$ should behaves as $\sqrt{\omega t_1}$.
Since interactions should be taken into account in this case, we should cautiously treat the HFE. At zero order in the HFE the effect of interactions is simple: it consists only in the initial one-site repulsion $\hat{H}_\text{eff}^{(0)} = \hat{H}_0 + \hat{H}_\text{BH}$.
At first order interactions do not contribute at all and the non-trivial effect of $\hat{H}_\text{BH}$ first appears only at second order in the HFE. These corrections  become irrelevant at constant $U$ provided $\omega \rightarrow \infty$. These questions and also the question of the relevance of the higher order terms in the HFE will be discussed in Sec.~\ref{sec:numFloquetCheck}.

\section{Engineering a bosonic FFLO analogue \label{sec:enginFmbec}}

\begin{figure}
\includegraphics[width=0.48\textwidth]
{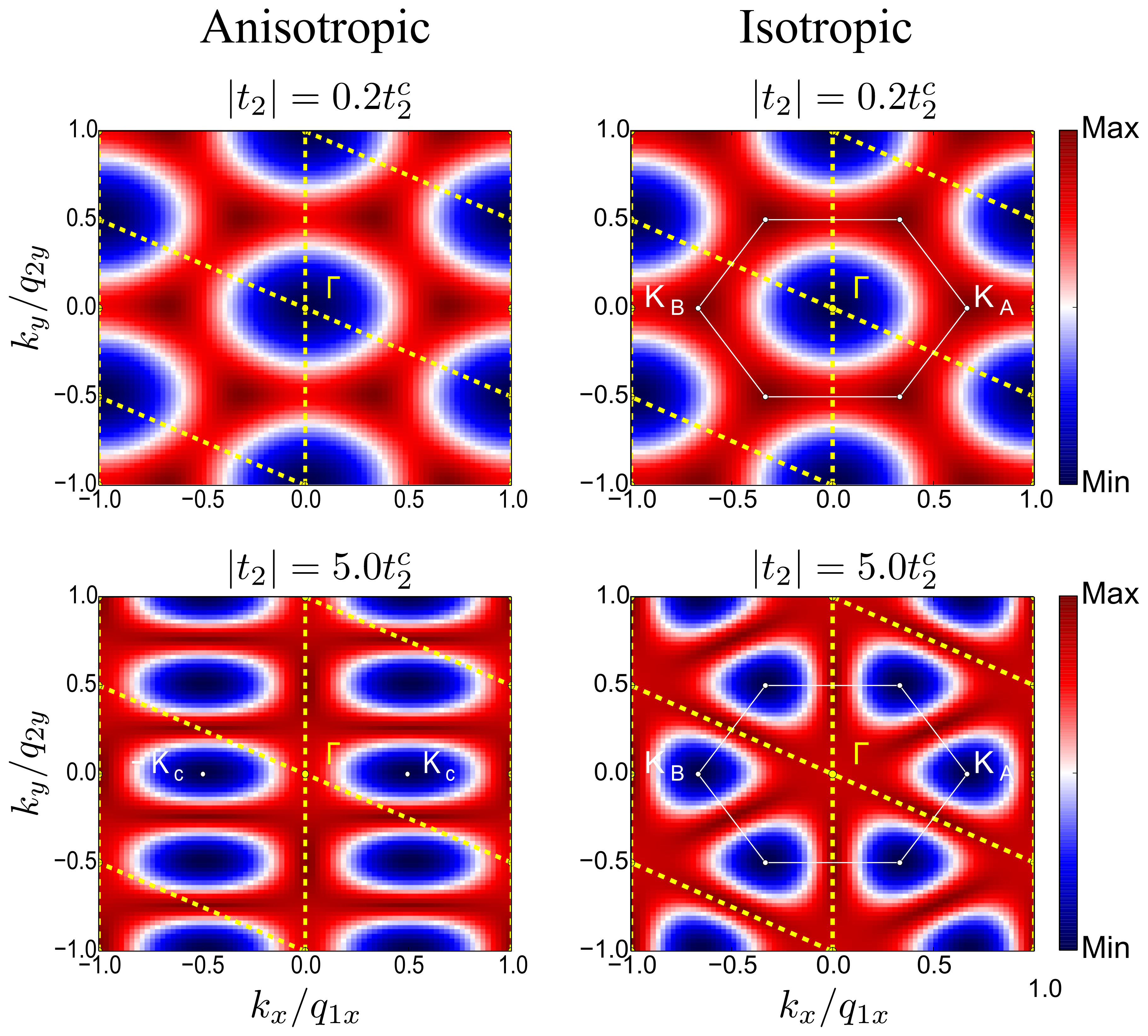}
\caption{Lowest energy band of the anisotropic Haldane model (on the left) and of the isotropic Haldane model (on the right) in the non-interacting case ($U=0$), plotted for different values of $t_2$ (different lines). Here $t_2^c$ is the "critical" value of the coupling when minima move away from the point $\bm{\Gamma}$. Yellow dotted lines represent reciprocal lattice generated by vectors $\bm{q}_1$ and $\bm{q}_2$ defined in Fig.~\ref{fig:anisHaldane_schemes}(d). Presence of artificial gauge fiels (copmlex NNN hoppings in our case) is required for observation of minima at points in the BZ different from the point $\bm{\Gamma}$.}
\label{fig:sfToCsfTransition}
\end{figure}

Following Ref.~\onlinecite{VassicPetrescu2015}, we explore the effect of weak interactions in the anisotropic Haldane model. 
Therefore, we consider the Hamiltonian
$\hat{H} = \hat{H}_0 + \hat{H}^{(1)}_\text{eff} + \hat{H}_{BH} $, where
\begin{align}
\label{eq:boseHubHam}
& \hat{H}_0 = 
	\epsilon_0 \Big( 
	\sum\limits_{i} \hat{a}^\dag_i \hat{a}_i +
	\sum\limits_{j} \hat{b}^\dag_j \hat{b}_j
	\Big) -
	t_1 \sum\limits_{\Braket{ij}}
	\left( 
        \hat{a}^\dag_i \hat{b}_j + \hat{b}^\dag_j \hat{a}_i
	\right) 
\notag \\
& \hat{H}^{(1)}_\text{eff} =
-t_2 \Big(
\sum\limits^\text{anis}_{\Braket{\Braket{ik}}}
	e^{\pm \frac{i\pi}{2}}\hat{a}^\dag_i \hat{a}_k + 
\sum\limits^\text{anis}_{\Braket{\Braket{jl}}}
	e^{\pm \frac{i\pi}{2}}\hat{b}^\dag_j \hat{b}_l + \text{h.c.}
\Big)
\notag \\
& \hat{H}_\text{BH} = 
\frac{U}{2} \sum\limits_{i} \hat{n}_i \left( \hat{n}_i - 1 \right)
\end{align}
For the sake of clarity, we choose $\omega_A = \omega_B = \epsilon_0$, such that the Semenoff mass term $M = \left(\omega_A-\omega_B\right)/2$, producing charge density wave orders in real space, is zero. When interactions are absent or sufficiently weak, this system is characterized by the presence of two distinct phases corresponding to the formation of either a zero-momentum BEC (\textit{ZM} phase) or a finite-momentum BEC (\textit{FM} phase) analogous to FFLO states. These phases appear since the single-particle Hamiltonian has minima at different points in the lowest band, as depicted in Figs.~\ref{fig:sfToCsfTransition} and \ref{fig:sfToCsfTansKy0}.
The \textit{ZM}--\textit{FM} transition between two phases is characterized by a change in the current patterns. We notice also that the nature of this transition is different in isotropic and anisotropic configurations of the Haldane model. In the first case the global minimum in the \textit{ZM} phase transforms into a local maximum in the \textit{FM} phase. Whereas in the second case the global minimum in the \textit{ZM} phase still stays a minimum in the \textit{FM} phase, but becomes local. This effect is due to the breaking of $C_3$ symmetry between two models. We outline also that considering a model with artificial gauge fields is the only possibility to observe the \textit{FM} phase using s-bands.

\subsection{ZM phase \label{sec:zmGs}}

The \textit{ZM} phase occurs as long as $t_2 < t_2^c = \sqrt{3/8} t_1$, while the isotropic model has a critical value $t_2^c = t_1/\sqrt{3}$. At zero temperature, bosons condense at zero momentum at the center $\bm{\Gamma}$ of the BZ, meaning that we can approximately write
\begin{align}
\hat{a}_i \approx \frac{\hat{a}_{\bm{\Gamma}}}{\sqrt{N_c}}	
, \qquad
\hat{b}_j \approx \frac{\hat{b}_{\bm{\Gamma}}}{\sqrt{N_c}}
\end{align}
Within this approximation, the Hamiltonian simplifies to
\begin{align}
\hat{H} &\approx
	\epsilon_0 \Big(
\hat{a}^\dag_{ \bm{\Gamma}} \hat{a}_{ \bm{\Gamma}} +
\hat{b}^\dag_{ \bm{\Gamma}} \hat{b}_{ \bm{\Gamma}}
\Big) - 
	3 t_1 \Big(
\hat{a}^\dag_{ \bm{\Gamma}} \hat{b}_{ \bm{\Gamma}} +
\hat{b}^\dag_{ \bm{\Gamma}} \hat{a}_{ \bm{\Gamma}}
\Big)
\notag \\
	& + \frac{U}{2} \Big[
\hat{a}^\dag_{ \bm{\Gamma}} \hat{a}_{ \bm{\Gamma}}
\Big( 
\frac{\hat{a}^\dag_{ \bm{\Gamma}} \hat{a}_{ \bm{\Gamma}}}{N_c} - 1
\Big) +
\hat{b}^\dag_{ \bm{\Gamma}} \hat{b}_{ \bm{\Gamma}}
\Big(
\frac{\hat{b}^\dag_{ \bm{\Gamma}} \hat{b}_{ \bm{\Gamma}}}{N_c} - 1
\Big) \Big]
\end{align}
We next introduce the total number of sites $N_s=2N_c$, the filling $n=N/N_s$ and the complex order parameter in sublattice A as $\Braket{\hat{a}_i} = \Braket{\hat{a}_{\bm{\Gamma}}}/\sqrt{N_c} = \sqrt{n} e^{i \theta_A}$, with a similar expression for sublattice B, assuming an equal filling in both sublattices.
The superfluid phases $\theta_{A/B}$ in each sublattice are pinned by the $t_1$ hopping-term, such that $\theta_A = \theta_B$. This corresponds to the presence of one Goldstone mode. The ground-state energy then reads
\begin{equation}
E^{ZM}_{GS} = N_s \left[ \left( \epsilon_0 - 3 t_1 \right) n + \frac{U}{2} n(n-1) \right]
\end{equation}
We remark that the contribution of $t_2$-terms effectively vanishes in the ground-state.
The bond currents are defined as
$ J_{ii'} = -2 \operatorname{Im}\left(
t_{ii'} \braket{\hat{a}^\dag_i \hat{a}_{i'}} \right) $ within sublattice A and similar expressions are used between sublattices A and B and within sublattice B. Here $t_{ii'}$ is the amplitude of the corresponding tunelling.
In the \textit{ZM} phase, currents between NNN sites in the direction of vectors 
$\bm{v}_1$ or $\bm{v}_2$ are proportional to $t_2$:
\begin{equation}
J^{ZM}_{AA, \bm{v}_1} =
J^{ZM}_{AA, \bm{v}_2} = -
2 \operatorname{Im}\left( -it_{2} n \right) = 2 n t_2 \;.
\end{equation}
In the anisotropic case we do not have NNN hoppings along $\bm{v}_3$, meaning that the corresponding currents are also always zero: $J^{ZM}_{AA, \bm{v}_3} = 0$.

\subsection{Excitations in the \textit{ZM} phase (Bogoliubov transformation) \label{sec:zmBglbv}}

In this section we use the Bogoliubov transformation \cite{Colpa1978BglbvTron} to study excitations above the groundstate in the \textit{ZM} phase.
When the temperature $T$ is sufficiently low, one can suppose that the ground state at $\bm{k} = 0$ will be still macroscopically occupied by a population of $N_0 = n_0 N_s$ bosons (with filling $n_0$). Small fluctuations can be described by operators $\hat{z}_{\nu, \bm{k}}$ with $\nu = a$ or $b$. 
The total number of particles is thus
\begin{equation}
N = N_0 + \sum\limits_{\substack{\bm{k} \in \mathcal{BZ}, \\ \bm{k} \neq 0, \nu}}
\hat{z}^\dag_{\nu, \bm{k}} \hat{z}_{\nu, \bm{k}}
\vspace{0.01pt}
\end{equation}
We are interested in writing the Hamiltonian $\hat{H}$ in powers of $\hat{z}_{\nu, \bm{k}}$. For convenience we also prefer to express $\hat{H}$ in the grand canonical ensemble, by indroducing the chemical potential $\mu$. To zero order in the perturbation we recognize the expression for the ground-state energy $E_{GS}^{ZM} - \mu N_0$.

The term linear in the fluctuation vanish for a particular value of the chemical potential, that is:
\begin{equation}
\mu = - 3t_1 + Un_0
\end{equation}
To second order in perturbation we obtain:
\\*
\begin{align}
\hat{H}_2 = &
\sum\limits_{\substack{\bm{k} \in \mathcal{BZ}, \\ \bm{k} \neq 0}}
\begin{bmatrix}
	\hat{z}^\dag_{a, \bm{k}} \\
	\hat{z}^\dag_{b, \bm{k}} \\
\end{bmatrix}^t \cdot \left[ -
\bm{d}(\bm{k}) \cdot \bm{\sigma} -
\mu I_{2\times2}
\right] \cdot
\begin{bmatrix}
	\hat{z}_{a, \bm{k}} \\
	\hat{z}_{b, \bm{k}} \\
\end{bmatrix}
\notag \\ & + 
	\frac{U n_0}{2} \sum\limits_{\substack{\bm{k} \in \mathcal{BZ}, \\ \bm{k} \neq 0, \nu}}
	\left[ 
		\hat{z}^\dag_{\nu, \bm{k}} \hat{z}^\dag_{\nu, -\bm{k}} +
		\hat{z}_{\nu, \bm{k}} \hat{z}_{\nu, -\bm{k}} +
		4 \hat{z}^\dag_{\nu, \bm{k}} \hat{z}_{\nu, \bm{k}}
	\right]
\end{align}
where functions $d_j(\bm{k})$ are defined in the Eq.~\eqref{eq:dFunctionsHaldane}.
The resulting Hamiltonian can now be rewritten (up to some constant term $E_0$) as
\begin{equation}
\hat{H}_2 = - \frac{1}{2}\sum\limits_{\bm{k}} 
\hat{Z}^\dag_{\bm{k}} \cdot 
\mathcal{H}_\text{PH} \left( {\bm{k}} \right) \cdot \hat{Z}_{\bm{k}}
\end{equation}
where $\hat{Z}_{\bm{k}}$ and $\mathcal{H}_\text{PH} \left( {\bm{k}} \right)$ 
are defined as follows:
\begin{widetext}
\begin{equation}
\hat{Z}_{\bm{k}}^\dag = 
\left[
	\hat{z}^\dag_{a,\bm{k}},
  	\hat{z}^\dag_{b,\bm{k}},
   	\hat{z}_{a,-\bm{k}},
  	\hat{z}_{b,-\bm{k}}
\right]
\end{equation}
\begin{equation}
\mathcal{H}_\text{PH} \left( {\bm{k}} \right) =
\begin{bmatrix}
  	\mu+d_z(\pmb{k})-2Un_0  &  d_x(\pmb{k})-d_y(\pmb{k}) i  &
  	-U n_0  &  0  \\
  	d_x(\pmb{k})+d_y(\pmb{k}) i  &  \mu-d_z(\pmb{k})-2Un_0  &
  	0  & -U n_0  \\
  	-U n_0  &  0 &
  	\mu-d_z(\pmb{k})-2Un_0 & d_x(\pmb{k})+d_y(\pmb{k}) i \\
  	0  &  -U n_0 &
  	d_x(\pmb{k})-d_y(\pmb{k}) i & \mu+d_z(\pmb{k})-2Un_0 \\
\end{bmatrix}
\end{equation}
\end{widetext}
This Hamiltonian can be diagonalized by using the Bogoliubov transformation from the particle operators $\hat{z}_{\nu, \bm{k}}$ to quasiparticle operators $\hat{\xi}_{\nu, \bm{k}}$. Transformed operators should preserve the bosonic commutation relations: 
\begin{align}
\left[ \hat{\xi}^\dag_{\nu_1, \bm{k}_1}, \hat{\xi}_{\nu_2, \bm{k}_2} \right] =& 
\delta_{\bm{k}_1 \bm{k}_2} \delta_{\nu_1 \nu_2} \\
\left[ \hat{\xi}_{\nu_1, \bm{k}_1}, \hat{\xi}_{\nu_2, \bm{k}_2} \right] =& 0
\end{align}

In the matrix notation this transformation can be written as $ \hat{Z}_{\bm{k}} = T_{\bm{k}} \cdot \hat{\Xi}_{\bm{k}} $. The condition of preserving commutation relations imposes restrictions on the form of $T_{\bm{k}}$. It should verify
\begin{equation}
\label{eq:BogoliubovTrans_ComRelCondition}
T_{\bm{k}} \cdot \Sigma \cdot T^\dag_{\bm{k}} =
T^\dag_{\bm{k}} \cdot \Sigma \cdot T_{\bm{k}} = 
\Sigma = 
\begin{bmatrix}
  	I_2  &  0   \\
  	0  &  -I_2  \\
\end{bmatrix}
\end{equation}
Our goal consists then in finding a transformation $T_{\bm{k}}$ that could diagonalize the Hamiltonian: 
$ T^\dag_{\bm{k}} \cdot \mathcal{H}_\text{PH} \left( {\bm{k}} \right) \cdot T_{\bm{k}} = 
\mathcal{D} \left( {\bm{k}} \right) $. If such a transformation exists in spite of constraints imposed by Eq.~\eqref{eq:BogoliubovTrans_ComRelCondition}, eigenvalues of the matrix $ \Sigma \mathcal{D} \left( {\bm{k}} \right)$ will correspond to the spectrum of Bogoliubov pseudo-particles.
One can then use the following identity: 
$ T^{-1}_{\bm{k}} \Sigma \mathcal{H}_\text{PH} \left( {\bm{k}} \right) T_{\bm{k}} = 
\Sigma \mathcal{D} \left( {\bm{k}} \right) $ 
to deduce that matrices $\Sigma \mathcal{H}_\text{PH} \left( {\bm{k}} \right)$ and $\Sigma \mathcal{D} \left( {\bm{k}} \right)$ are similar. Thus, in order to obtain the Bogoliubov spectrum we only need to diagonalize $\Sigma \mathcal{H}_\text{PH} \left( {\bm{k}} \right)$.

\begin{figure}
\includegraphics[width=0.48\textwidth]
{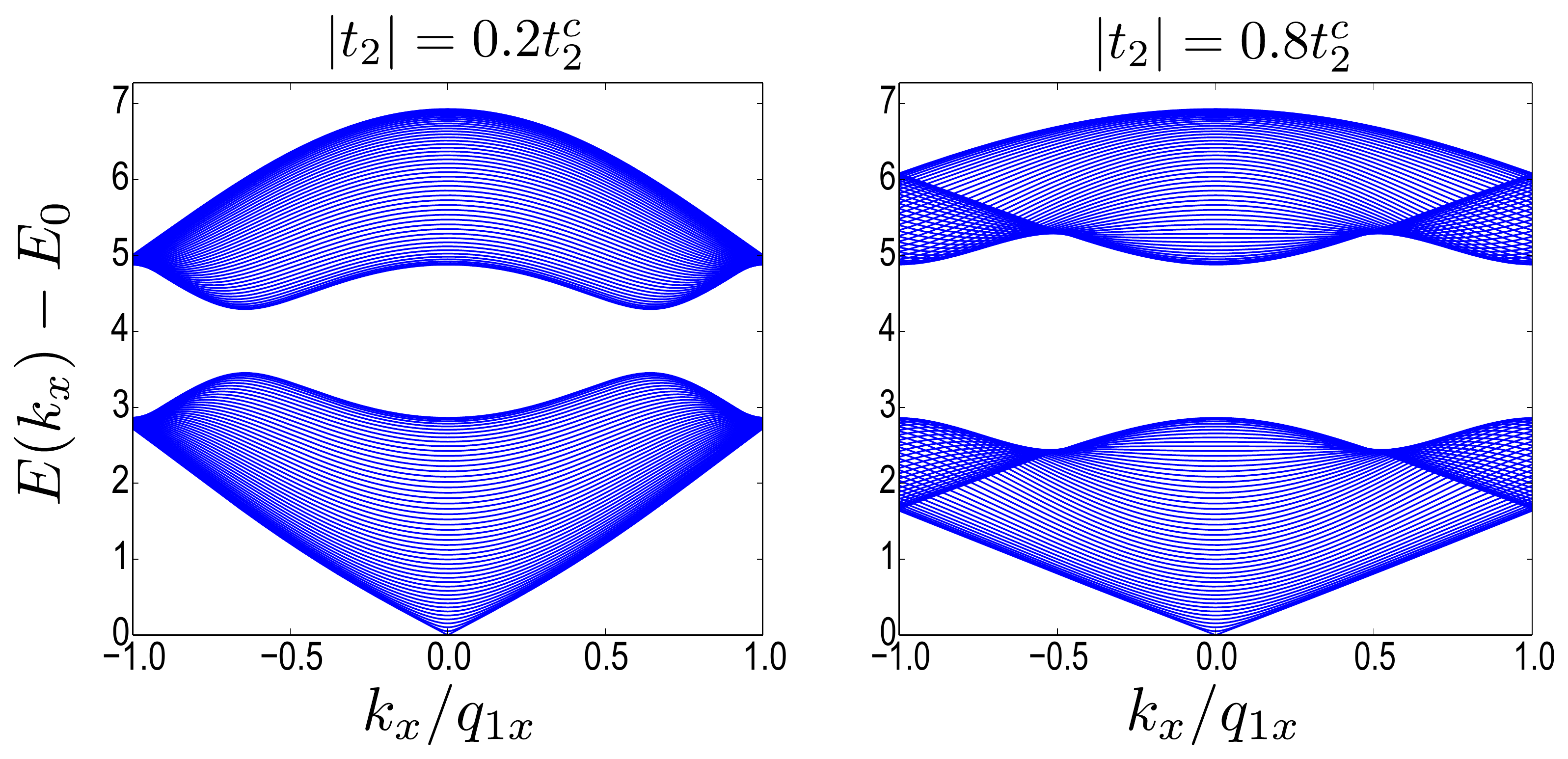}
\caption{Dispersion relation of excitations in the \textit{ZM} phase of the anisotropic Haldane model, showing the linear Bogoliubov dispersion relation around the point $\bm{\Gamma}$ in the BZ. All figures were taken for $t_1 = 1$ and $n_0 U = 1$.}
\label{fig:bglbv_2D_SF}
\end{figure}

In Fig. \ref{fig:bglbv_2D_SF} we show the corresponding solution for the anisotropic Haldane-like model at the value of the interaction such that $n_0 U = t_1 = 1$. The dispersion relation is linear around the point $\bm{k} = 0$. We remark that in the anisotropic case the velocity of the "sound" mode depends on $t_2$, whereas in the isotropic case it was completely independent of $t_2$. In Ref.~\onlinecite{Furukawa2015topoBogoliubonsHaldaneIs} it was shown for the case of the isotropic Haldane model, that topological properties of Bloch bands present in the noninteracting case are smoothly carried over to Bogoliubov excitation bands in the \textit{ZM} phase. We expect the same formalism to be applicable to the case of the anisotropic Haldane model and in the \textit{FM} phase.

\subsection{Critical value of the NNN hopping amplitude \label{sec:sfToCsf_t2crit}}

\begin{figure*}
\includegraphics[width=1.0\textwidth]
{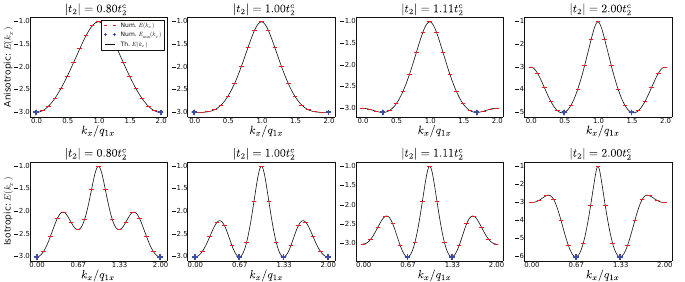}
\caption{Cut of the lowest energy band of the single-particle system
along the $k_y = 0$ line.
Analytical calculations (black solid line) are superposed 
to the results of ED (red dashes).
\textbf{(Upper panel)} Anisotropic model. Structure of the ED cluster is 
$10 \times 2$ cells (40 sites) with tilt $t_h = -1$.
Values of $t_2$ are chosen is such a way that the condensation occurs precisely
at the points allowed by the geometry of the ED cluster.
The value $t_2 = 2 t_2^c$ is high enough to see the physics of the $t_2 \gg t_2^c$ regime. In this last case, shown on the rightmost figure, only the small distinction
between analytical and numerical results (due to finite size effects) is observed. 
\textbf{(Lower panel)} Isotropic model. Structure of the ED cluster is 
$9 \times 2$ cells (36 sites) with tilt $t_h = -1$.}
\label{fig:sfToCsfTansKy0}
\end{figure*}

As we increase the ratio $t_2/t_1$ up to some critical value, the minimum of the single-particle band structure at the point $\bm{\Gamma}$ splits into two new minima that then start moving away from the center of the BZ. Numerical simulations of the Fig. \ref{fig:sfToCsfTransition} and symmetry arguments imply that for any value of $t_2/t_1$, position of minima along the $y$-axis does not change. Thus, it should be sufficient for us to perform calculations only along the $k_y = 0$ axis in the BZ. In the Fig. \ref{fig:sfToCsfTansKy0} we show some examples of the analytical energy structure compared to results of ED simulations for the 1-particle Hamiltonian in different regimes of $t_2$ for $k_y = 0$, $k_x \in \left[ 0,4\pi/\sqrt{3}a \right]$. 

The goal of this subsection is to deduce the precise value of the critical NNN hopping amplitude $t_2^c$. Let us consider that the condensation in the general case occurs at points $\bm{k} = \pm \left( z\pi/\sqrt{3}a \right) \bm{e}_x$, for some real $z \in [0, 2]$, defined up to a reciprocal lattice vector. Eigenvalues of the 1-particle Hamiltonian $\hat{H}_{1p}$ at these points are
\begin{equation}
\epsilon_\pm(z) = \epsilon_0 \pm \sqrt{5 t_1^2 + 
\left( 16 t_2^2 - 4t_1^2 \right) \sin^2 \left( \frac{\pi z}{2} \right) +
4 t_1^2 \cos \left( \frac{\pi z}{2} \right)}
\end{equation}
Extrema of the lower band energy correspond to zeros of $\frac{\partial \epsilon_-(z)}{\partial z}$, that we will denote by $z_c$. They are solutions of the following set of equations:
\begin{equation}
\left[
\begin{array}{ccc}
\sin(\frac{\pi z_{c}}{2}) & = & 0 \\
\cos(\frac{\pi z_{c}}{2}) & = & \frac{1}{2}
\left( \frac{t_1^2}{4t_2^2 - t_1^2} \right)
\end{array}
\right.
\end{equation}
We deduce three main regimes in the evolution of the energy band structure along the $k_y = 0$ line. 

The first regime is determined by $\left| t_2 \right| < \sqrt{1/8} t_1$. In this region the solution of the second equation exists and corresponds to the maximum of the band at two Dirac points. If we increase $t_2$, starting from zero, the value of $z_c$ corresponding to these maxima increases (Dirac points move away from extremities of the BZ) until it reaches the value $z_c = 2$. The minimum of the band is localized at the point $\bm{\Gamma}$, that corresponds to the solution of the first equation for $z_c = 0$.

In the region $\sqrt{1/8} t_1 < \left| t_2 \right| < \sqrt{3/8} t_1$
there is no solution of the second equation and the band possess only one minimum at $z_c = 0$ and maximum at $z_c = 2$.

In the region $\left| t_2 \right| > \sqrt{3/8} t_1$ the solution of the second equation starts existing again and corresponds to two new minima of the band. At the same time, the solution of the first equation at $z_c = 0$ transforms into a local maximum and the solution at $z_c = 2$ stays a global maximum of the band. This defines the precise point of the \textit{ZM} -- \textit{FM} transition: $t^c_2 = \sqrt{3/8} t_1$.

Transition between all these three regimes can be clearly seen in the Fig. \ref{fig:sfToCsfTansKy0}(upper panel). The shape of the energy band at the critical point, expressed in terms of the parameter $z$ is
\begin{equation}
\epsilon_-(z) \underset{z \rightarrow 0}{\approx}
\epsilon_0 - 3t_1 + 
\frac{t_1}{12} \left( \frac{\pi z}{2} \right)^4
\end{equation}
Thus, at the transition the well becomes much wider: the second power in the dispersion relation is replaced by the fourth power. In the following we will study more closely the behaviour of the time-dependent system in the regime of the effective NNN hopping amplitude $t_2$ close to $t_2^c$. However, in order to give an estimation for required experimental parameters, we calculate the critical value $V_c$ needed to reach the \textit{ZM} -- \textit{FM} phase transition at a fixed frequency: $V_c (\omega = 60 t_1) \approx 8.57 t_1$.

Displacement of the condensation point for different parameters of the problem affects the procedure of the numerical ED. Numerical simulations of the infinite lattice requires considering a particular set of points on the lattice, forming a cluster, and of periodic boundary conditions gluing together different clusters. This defines the set of quantum numbers associated to irreducible representations of the group of translations along two Bravais vectors of the lattice. Each quantum number corresponds to one point in the BZ and the amount of quantum numbers equals the number of possible translations in the cluster. Thus, by increasing the cluster size, we increase the resolution in the BZ.

\begin{figure}
\includegraphics[width=0.48\textwidth]
{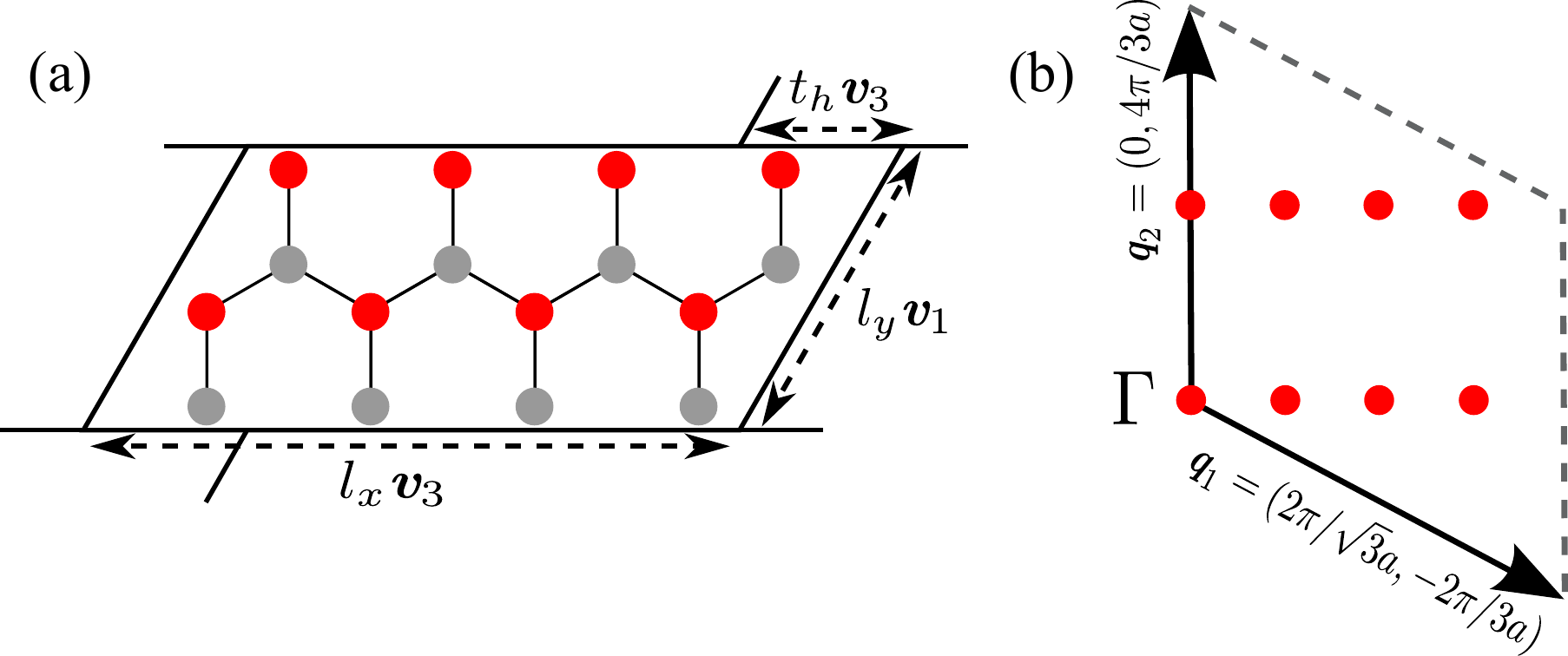}
\caption{\textbf{(a)} Schematic representation of the tilted cluster with $4 \times 2$ cells and with the tilt $t_h = -1$, used in ED simulations of the \textit{ZM} -- \textit{FM} phase transition. Black lines determine boundary of each cluster.
\textbf{(b)} Corresponding quantum numbers allowed by symmetries of the cluster.}
\label{fig:tilt_lattice}
\end{figure}

According to our previous discussion, in order to capture the \textit{ZM} -- \textit{FM} transition in the anisotropic model, one can restrict ourselves to the case $k_y = 0$. This motivates us to consider a particular choice of clusters, maximizing the resolution in the BZ along $k_x$ with the price of minimizing the resolution along $k_y$. We make such clusters by composing $l_x \times l_y$ unit cells of the honeycomb lattice in the geometry of the parallelogram generated by vectors $l_x \bm{v}_3$ and $l_y \bm{v}_1$. Each parallelogram is then translated with spanning vectors $\bm{s}_1 = l_x \bm{v}_3$ and $\bm{s}_2 = l_y \bm{v}_1 - t_h \bm{v}_3$ to fill the full lattice. Here $t_h$ denotes the tilt. This defines our implementation of PBC. Example of such tilted cluster is shown in Fig.~\ref{fig:tilt_lattice}.

\subsection{FM phase \label{sec:fmPhase}}

The \textit{FM} phase arises when $t_2$ reaches the critical value $t_2^c = \sqrt{3/8} t_1$. In the limit $t_2 \gg t_1$, the two sublattices become decoupled and bosons condense at the two inequivalent points
$ \bm{K}_A = \left( \pi/\sqrt{3} a \right) \bm{e}_x $ (on sublattice $A$), and 
$ \bm{K}_B =-\left( \pi/\sqrt{3} a \right) \bm{e}_x $ (on sublattice $B$), as shown on Fig.~\ref{fig:sfToCsfTransition}(c). We begin with the study of the properties of the finite momentum BEC, by considering first this simpler case. 

\subsubsection{FM phase in the regime of decoupled sublattices }

In the limit of two decoupled sublattices, one can use the following approximation for the real space annihilation (and creation) operators:
\begin{align}
\hat{a}_i \approx
	\frac{e^{-i \bm{r}_i \bm{K}_A}}{\sqrt{N_c}}	
	\hat{a}_{\bm{K}_A}
, \qquad
\hat{b}_j \approx
	\frac{e^{-i \bm{r}_j \bm{K}_B}}{\sqrt{N_c}}
	\hat{b}_{\bm{K}_B}
\end{align}
In the following we will also use the notation
\begin{align}
\braket{\hat{a}^\dag_i\hat{a}_i} =
n_A = N_A/N_c
, \qquad
\braket{\hat{b}^\dag_j \hat{b}_j} = n_B = N_B/N_c
\end{align}
such that $\left( n_A + n_B \right)/2 = n = N/N_s$ is the filling, since $N_c=N_s/2$.
Within the mean-field approximation, we simply write
\begin{equation}
\braket{\hat{a}_{\bm{K}_A}} \approx \sqrt{N_A}e^{i\theta_A}
, \qquad
\braket{\hat{b}_{\bm{K}_B}} \approx \sqrt{N_B}e^{i\theta_B}\;,
\end{equation}
where $\theta_A$ and $\theta_B$ are the superfluid phases associated with the two condensates.
One can easily check that in the groundstate the contribution of the NN hopping term vanishes in the thermodynamic limit:
\begin{align}
\sum\limits_{\braket{ij}} \hat{a}^\dag_i \hat{b}_j & 
\propto
\sum\limits_i \left( e^{-i\bm{K}_B\textbf{r}_i} 
	\sum\limits_{j = 1}^3 
	 e^{-i\bm{K}_B\textbf{u}_j}
	\hat{a}^\dag_i \hat{b}_{\bm{K}_B} \right)
\notag \\ &= 
\sum\limits_i 
e^{i \left( \bm{K}_A - \bm{K}_B \right) \bm{r}_i}
\hat{a}^\dag_{\bm{K}_A} \hat{b}_{\bm{K}_B} = 0\;.
\end{align}
We thus see that the contribution of the $t_1$ term effectively vanishes in the groundstate.
Finally, we obtain that in the regime $t_2 \gg t_2^c$, the mean-field approximation simplifies the Hamiltonian into
\begin{align}
\hat{H} \approx &
	\left( \epsilon_0 - 4 t_2 - U/2\right)
\left(
\hat{a}^\dag_{ \bm{K}_A} \hat{a}_{ \bm{K}_A} +
\hat{b}^\dag_{ \bm{K}_B} \hat{b}_{ \bm{K}_B}
\right)
\notag \\ 
& + \frac{U}{N_s} \left[
\left( 
\hat{a}^\dag_{ \bm{K}_A} \hat{a}_{\bm{K}_A}
\right)^2 +
\left(
\hat{b}^\dag_{ \bm{K}_B} \hat{b}_{ \bm{K}_B}
\right)^2 \right]
\end{align}
so that the groundstate energy reads
\begin{equation}
E^{FM}_{GS} \underset{t_2 \gg t_2^c}{=} N_s \left[
\left( \epsilon_0 - 4 t_2 - U/2 \right) n + 
U \left(n_A^2+n_B^2\right)/4 \right]\;.
\end{equation}
As the two sublattices are completely decoupled, both phases $\theta_A$ and $\theta_B$ become independent parameters, which corresponds to the presence of two Goldstone modes. 

The phase difference $\theta_A - \theta_B$ can be fixed if a coherent coupling between the two wells is generated, resulting in a modification of $J_{AA}$ and allowing for finite inter-sublattice currents $J_{AB}$. Such effect can be induced, for example, by adding impurities to the model or, in the thermodynamic limit, by taking into account quantum fluctuations via the so called "order by disorder" mechanism \cite{Henley1989orderByDisorder}. This problem was studied in details in Ref.~\onlinecite{VassicPetrescu2015} in the case of the isotropic Haldane model, and it will not be considered in this work. In numerical simulations, we also do not observe this effect since the groundstate obtained using ED corresponds to the state without coherent coupling. In particular, NN currents $J_{AB}$ calculated in the groundstate using ED will always be zero. As a consequence, in the following we will refer to the relatively simple case of two condensates at $\bm{K}_A$ and $\bm{K}_B$, that are completely decoupled. This leads in particular to the conclusion that in the $t_2 \gg t^c_2$ regime of the \textit{FM} phase NNN currents are simply zero:
\begin{equation}
J^{FM}_{AA, \bm{v}_j} \underset{t_2 \gg t_2^c}{=} 0\;.
\end{equation}

\subsubsection{FM phase in the intermediate regime \label{sec:csfPhase}}

\begin{figure}
\centering
\includegraphics[width=0.4\textwidth]
{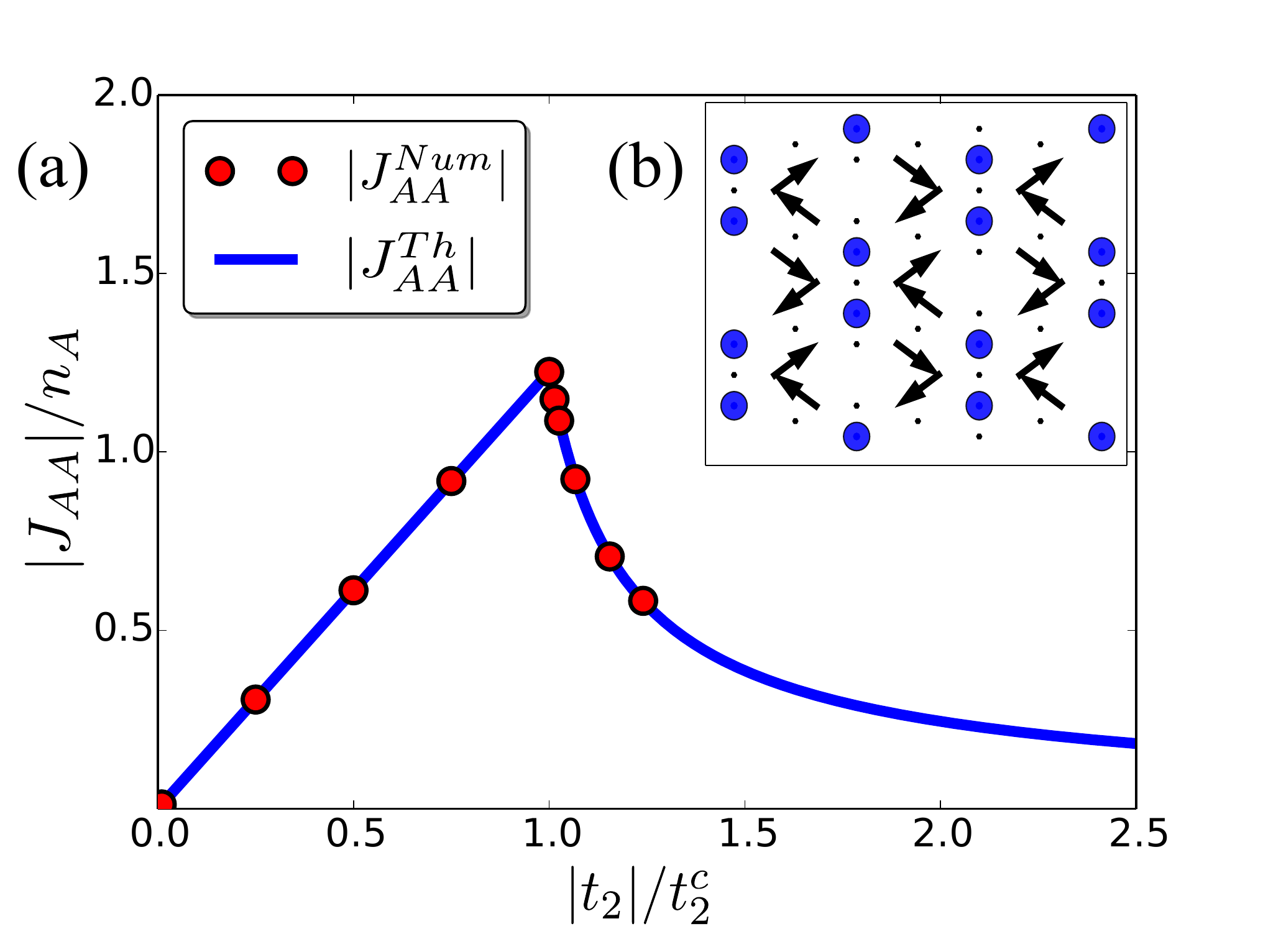}
\caption{\textbf{(a)} Variation of the second neighbour current $J_{AA}$ with $t_2$.
Qualitative change of behaviour can be seen when crossing the critical value $t_2^c = \sqrt{3/8}t_1$. In the figure analytical prediction (blue line) is compared with the results of ED (red circles). \textbf{(b)} Result of ED showing patterns of currents $J_{AA}$. Closed loops of chiral currents involve lozenge geometries. Different types of arrows correspond to different sublattices.}
\label{fig:jAaCsf}
\end{figure}

We now consider the more realistic case when $t_2$ becomes of the order of $t_1$ in the $t_2 > t_2^c$ regime. We then define  $\hat{\alpha}_{\bm{k}}$ and $\hat{\beta}_{\bm{k}}$ the annihilation operators in the lower and upper energy bands of the single-particle Hamiltonian. At each point of the BZ these operators are related to $\hat{a}_{\bm{k}}$ and $\hat{b}_{\bm{k}}$ through some unitary transformation:
\begin{align}
\label{eq:unitaryTransformation}
\hat{a}_{\bm{k}} =& 
	\mu(\bm{k}) \hat{\alpha}_{\bm{k}} +
	\nu(\bm{k}) \hat{\beta}_{\bm{k}} 
\notag \\
\hat{b}_{\bm{k}} =&
	e^{i\phi(\bm{k})} \left[ -
	\nu^*(\bm{k}) \hat{\alpha}_{\bm{k}} +
    \mu^*(\bm{k}) \hat{\beta}_{\bm{k}} \right]
\end{align}
with $|\mu(\bm{k})|^2 + |\nu(\bm{k})|^2 = 1 $. 
In the low temperature limit, bosons condense at points $\pm \bm{K}_c = \pm \left( z_c \pi/\sqrt{3}a \right) \bm{e}_x$ (we recall that by our definition $z_c > 0$) so that 
$\braket{\hat{\alpha}_{\pm \bm{K}_c}} \sim \sqrt{N}$ and 
$\braket{\hat{\beta}_{\pm \bm{K}_c}} = 0$.
Thus, annihilation (and creation) operators in the real space are approximated by
\begin{align}
\label{eq:csfGs}
\hat{a}_i &\approx \frac{
	e^{-i \bm{K}_c \bm{r}_i} \hat{a}_{ \bm{K}_c} +
	e^{ i \bm{K}_c \bm{r}_i} \hat{a}_{-\bm{K}_c}
	}{\sqrt{N_c}}
\notag \\ 
\hat{b}_i &\approx \frac{
	e^{-i \bm{K}_c \bm{r}_i} \hat{b}_{ \bm{K}_c} +
	e^{ i \bm{K}_c \bm{r}_i} \hat{b}_{-\bm{K}_c}
	}{\sqrt{N_c}}\;.
\end{align}
We then introduce the averages 
\begin{equation}
\braket{\hat{a}_{ \pm \bm{K}_c}} =
\sqrt{N_{A,\pm}}e^{i\theta_{A,\pm}}
, \qquad
\braket{\hat{b}_{ \pm \bm{K}_c}} =
\sqrt{N_{B,\pm}}e^{i\theta_{B,\pm}}
\end{equation}
such that $N_{A,+} + N_{A,-} = N_A $, $N_{B,+} + N_{B,-} = N_B$,
$N_{A,\pm} + N_{B,\pm} = N_\pm $ and $N_A + N_B = N_+ + N_- = N$.
The approximation \eqref{eq:csfGs} can be used to find a more general form of the GS Hamiltonian and GS energy in the \textit{FM} phase (see Appendix \ref{app:fmPhaseInter} for their complete expressions). Moreover, we obtain that phases $\theta_{A,\pm}$ and $\theta_{B,\pm}$ are be pinned pair by pairs, which corresponds to the presence of two Goldstone modes. We also get the expression of currents valid in both regimes of $t_2$.
\begin{align}
\label{eq:fmbecCurrentTh}
J_{AA, \bm{v}_1} = J_{AA, \bm{v}_2} =&
-t_2\frac{N_A}{N_c}
\cos \left( z_c \frac{\pi}{2} \right)
\notag \\
J_{AA, \bm{v}_3} =& 0
\end{align}
The evolution of the currents $J_{AA}$ with $t_2$ is shown on Fig.~\ref{fig:jAaCsf}, where the analytical prediction is compared with the results of ED in a weakly interacting regime.

\begin{figure}
\centering
\includegraphics[width=0.4\textwidth]
{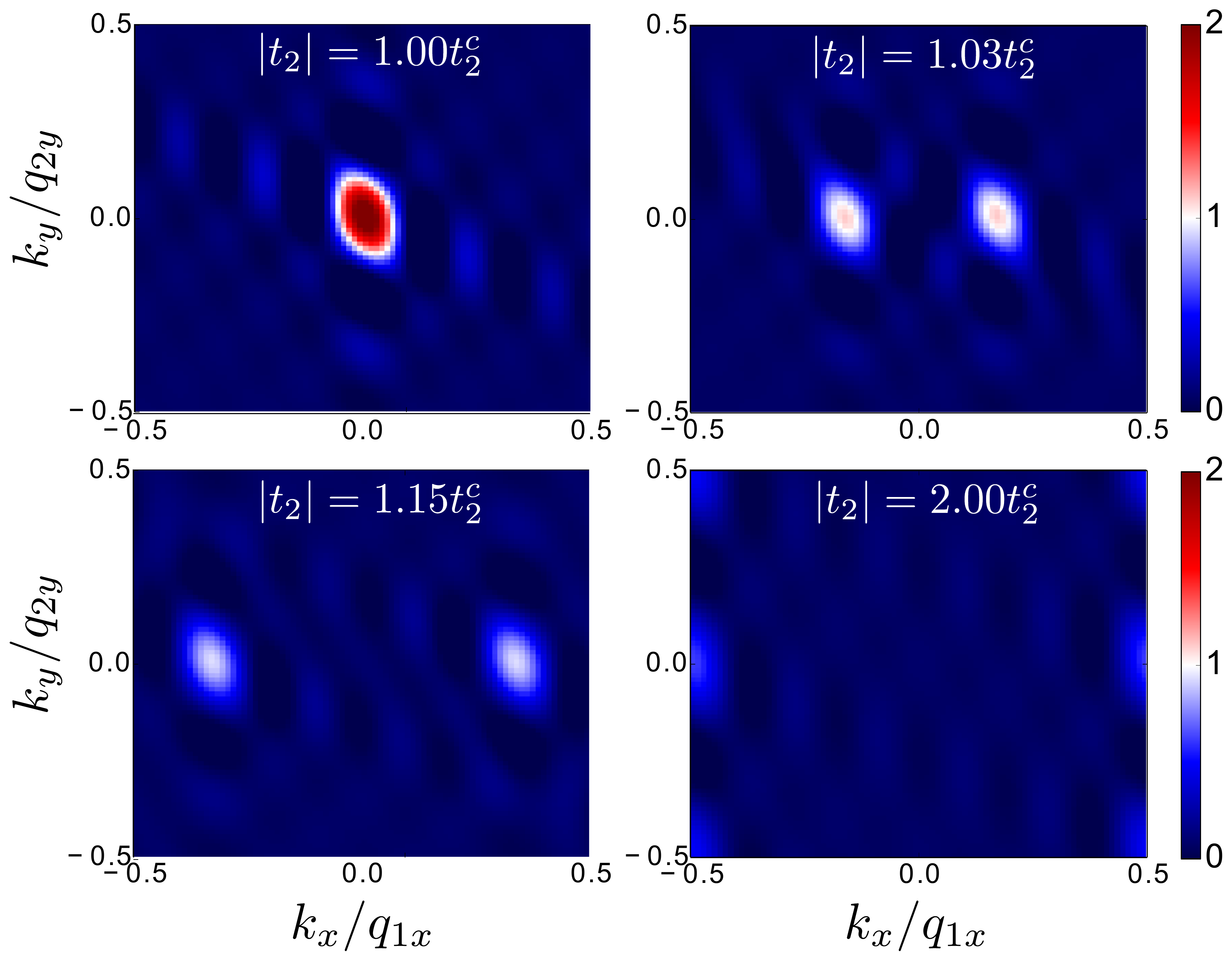}
\caption{Momentum distribution $n(\bm{k})$ at the transition in the \textit{FM} phase in the bosonic many-body ground-state. Structure of the ED cluster is $6\times4$ cells (48 sites) with tilt $t_h = -2$. Parameters of simulations: $N = 2$, $U = 0$. Finite size effects result in a tilt of the maxima of $n(\bm{k})$ and its non-physical oscillations.}
\label{fig:nkCsf}
\end{figure}

Another way to perform a numerical verification of the behavior of the system in the \textit{FM} regime is to look at the momentum distribution $n(\bm{k})$. In Fig.~\ref{fig:nkCsf}, we give some results of ED showing how the momentum distribution in the many-body GS in the $\mathbf{k}=0$ sector evolves when changing $t_2$ above the critical value $t_2^c$. We effectively observe the splitting of the condensation point into two points and their further displacement along the $k_y = 0$ line.

The continuous variation of currents suggests that the \textit{ZM}--\textit{FM} transition in the anisotropic Haldane model is of the second order. In order to justify this point, we first write explicitly the unitary transformation of Eq.~\eqref{eq:unitaryTransformation} at momentum $\pm\bm{K}_c$ in terms of $z_c$:
\begin{align}
\label{eq:unitaryTransformationZc}
	\hat{a}_{\pm\bm{K}_c} = &-
\sqrt{\frac{\sqrt{X^2(z_c)+Y^2(z_c)}\pm Y(z_c)}{2\sqrt{X^2(z_c)+Y^2(z_c)}}}
\hat{\alpha}_{\pm\bm{K}_c}
\notag \\ &+ 
\sqrt{\frac{\sqrt{X^2(z_c)+Y^2(z_c)}\mp Y(z_c)}{2\sqrt{X^2(z_c)+Y^2(z_c)}}}
\hat{\beta}_{\pm\bm{K}_c}
\notag \\
	\hat{b}_{\pm\bm{K}_c} = &-
\sqrt{\frac{\sqrt{X^2(z_c)+Y^2(z_c)}\mp Y(z_c)}{2\sqrt{X^2(z_c)+Y^2(z_c)}}}
\hat{\alpha}_{\pm\bm{K}_c} 
\notag \\ &-
\sqrt{\frac{\sqrt{X^2(z_c)+Y^2(z_c)}\pm Y(z_c)}{2\sqrt{X^2(z_c)+Y^2(z_c)}}}
\hat{\beta}_{\pm\bm{K}_c} 
\end{align}
where $X(z_c)$ and $Y(z_c)$ are defined as follows:
\begin{align}
X(z_c) &= t_1 \left[ 1 + 2 \cos\left( z_c \frac{\pi}{2} \right) \right]
\notag \\
Y(z_c) &= 4 t_2 \sin\left( z_c \frac{\pi}{2} \right)
\end{align}
One notices that $N_+$ and $N_-$, corresponding to the number of particles in the wells at $\pm\bm{K}_c$, and $z_c$ are the only parameters of the problem. In Eq.~\eqref{eq:eGsCsfIntEffects}, we express the GS energy in terms of these quantities. 
Close to the \textit{ZM}--\textit{FM} phase transition, we expand the GS energy in powers of $z_c$. By doing this calculation in the non-interacting case we obtain
\begin{align}
& E_{GS} \underset{z_c \rightarrow 0}{=}
\text{cst} -
t_1 N
\left[ \frac{8}{3}
\left( \frac{t_2}{t_1} \right)^2 - 1 \right]
\left( \frac{\pi z_c}{2} \right)^2
\notag \\ & +
\frac{N}{108t_1^3}
\left( 128 t_2^4 - 9t_1^4 \right)
\left( \frac{\pi z_c}{2} \right)^4 + \dots
\end{align}
The sign change of the first term, proportional to $z_c^2$, occurs at the value $t_2^c = \sqrt{3/8}t_1$, in agreement with our previous estimations. Moreover, the sign of the second term, proportional to $z_c^4$, is always positive in the \textit{FM} phase. Thus, we clearly see that the transition in the anisotropic Haldane model is of the second order. This is different from the case of the isotropic Haldane model \cite{VassicPetrescu2015, Kuno2015bosonicIsHaldane}, where the transition is of the first order, since three wells are simultaneously present at the transition.

\subsection{Effect of interactions on the QPT \label{sec:sfToCsf_ints}}

\begin{figure*}
\includegraphics[width=1.0\textwidth]
{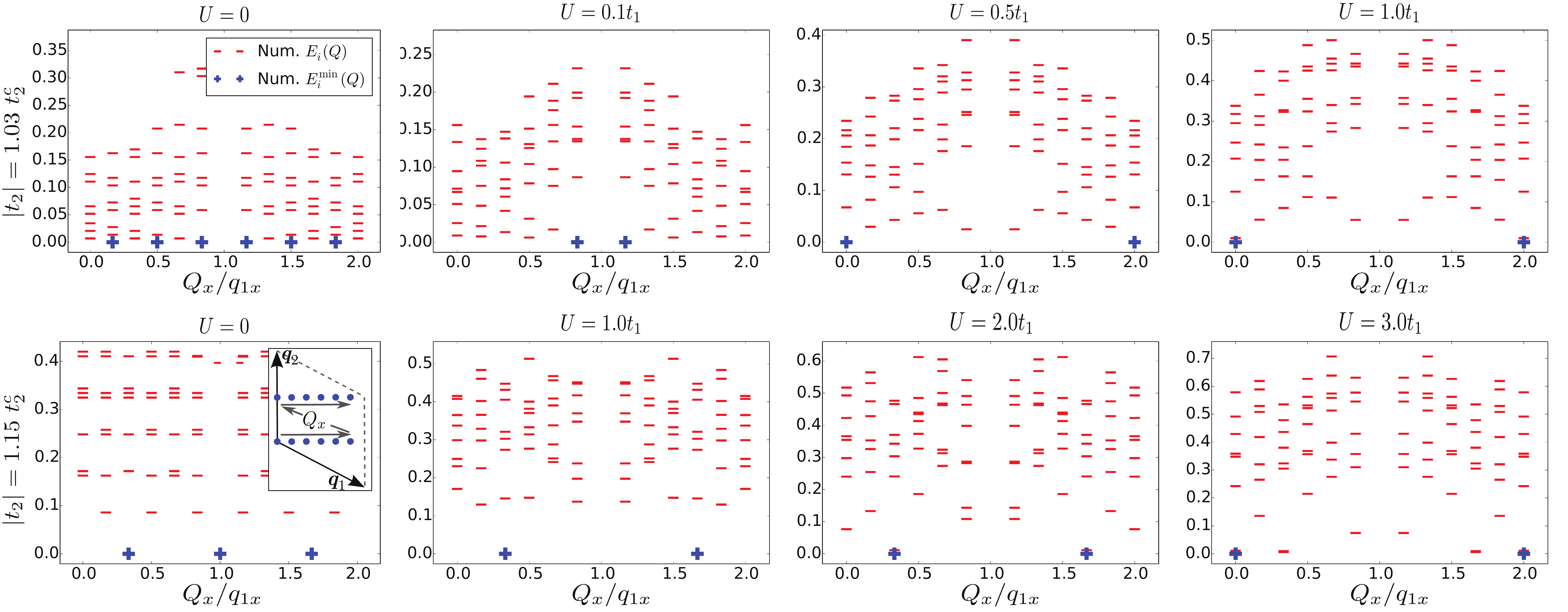}
\caption{Many-body energy levels (cut along the $k_y = 0$ line in the BZ). ED cluster: $6\times2$ cells (24 sites). Position in the BZ of quantum number associated to this precise cluster are shown on a small replica of Fig.~\ref{fig:tilt_lattice}(b). Different columns correspond to different values of $U$. Different lines correspond to different values of $t_2$. Parameters: $N=5$ particles.}
\label{fig:qnsEnsCsfWithUN5}
\end{figure*}
\begin{figure*}
\includegraphics[width=1.0\textwidth]
{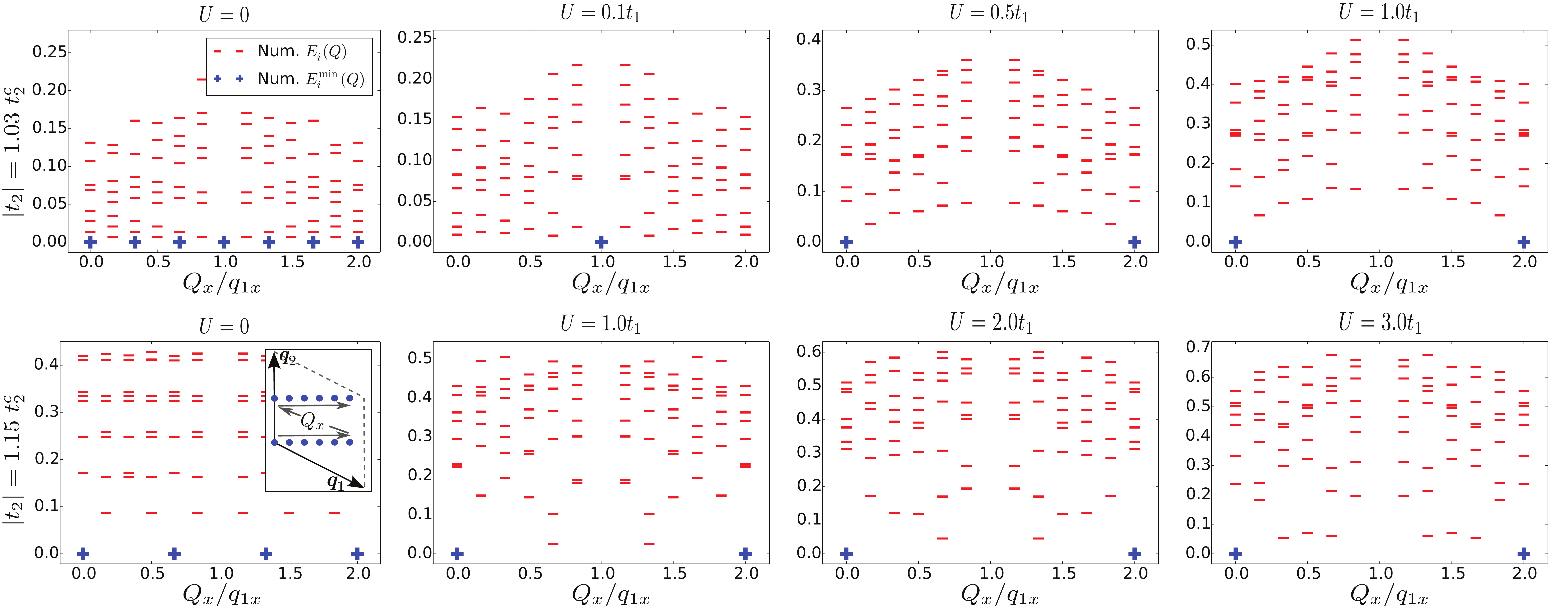}
\caption{Many-body energy levels (cut along the $k_y = 0$ line in the BZ). Energy minima are shown in blue. ED cluster: $6\times2$ cells (24 sites), the tilt $t_h = -1$. Position in the BZ of quantum number associated to this precise cluster are shown on a small replica of Fig.~\ref{fig:tilt_lattice}(b). Different columns correspond to different values of $U$. Different lines correspond to different values of $t_2$. Parameters: $N=6$ particles.}
\label{fig:qnsEnsCsfWithUN6}
\end{figure*}

In order to study more precisely the effect of interactions on the \textit{ZM} -- \textit{FM} phase transition, we repeat the previous analysis and write down the contribution to the GS energy in powers of $z_c$ associated with the interaction term only. This calculation is based on the ansatz \eqref{eq:csfGs}. Thus, we must assume that interactions are weak enough so that they do not affect the two-well structure of the system in the \textit{FM} phase.
The result of such a calculation is given in Eq.~\eqref{eq:eGsCsfIntEffects} of Appendix~\ref{app:fmPhaseInter}.

First, we see that in the \textit{FM} phase interactions imposes constraints onto $N_+$ and $N_-$ -- the numbers of particles in each well, and removes the degeneracy that was present in the non-interacting system. However, contrary to the most intuitive guess, there are two groundstates that minimize the energy in two distinct sub-regimes of the \textit{FM} phase: for $t_2^c < t_2 < \sqrt{(17+\sqrt{97})/24}\ t_2^c$ interactions favor particles in one particular well, such that either $N_+$ or $N_-$ becomes precisely equal to $N$. When $t_2 > \sqrt{(17+\sqrt{97})/24}\ t_2^c$ the uniform distribution of particles $N_+ = N_- = N/2$ is preferred.

The second effect of weak interactions consists in moving the position of minima away from $\pm \bm{K}_c$. In particular, to the lowest order in $z_c$, the contribution of interactions to the GS energy at the \textit{SF}--\textit{CSF} transition is
\begin{align}
\Delta E'_{GS} = NUn\frac{8 t_2^2}{9t_1^2} \left(z_c\frac{\pi}{2}\right)^2
\end{align}
This leads to the increase of the critical NNN coupling amplitude $t_2^c$.
If particles are not located all in one well, there appears an additional contribution
\begin{align}
\Delta E''_{GS} = -
\frac{16U}{3}
\frac{N_+\left(N-N_+\right)}{N_s}
\left(\frac{t_2}{t_1}\right)^2
\left(z_c\frac{\pi}{2}\right)^2
\end{align}
that, on the opposite, tends to decrease $t_2^c$. The second contribution $\Delta E''_{GS}$ dominates when $N_+ = N_- = N/2$, resulting in the generation of an effective repulsion between the two wells (see also Appendix~\ref{app:fmPhaseInter} for a more detailed analysis).

In order to check our theoretical predictions, we perform ED calculations. We plot the many-body energy levels $E_i(k_x)-E_0$ for different values of the total momentum $\bm{k}$ (here $E_0$ denotes the lowest energy over all momentum sectors). Simulation results are shown in Figs.~\ref{fig:qnsEnsCsfWithUN5} for an odd number of particles and \ref{fig:qnsEnsCsfWithUN6} for an even number of particles.

Without interactions, we observe a high degeneracy of the many-body groundstate due to the fact that all bosons condense independently in one of the two wells. 
If one increases the interaction strength, this degeneracy is lifted. We observe that, close to the transition ($t_2 \approx t_2^c$) in the \textit{FM} phase, the most energetically favored groundstate of the weakly interacting regime corresponds to the state with all particles condensed in the same well. This is not the case for higher values of $t_2$. If $t_2$ increases, the groundstate becomes, as expected, the state with all particles uniformly distributed over two wells (with $N_+ - N_- = \pm 1$ if $N$ is odd). The value at which this transition occurs agrees well with the analytical prediction $t_2 = \sqrt{(17+\sqrt{97})/24}\ t_2^c$, with some imprecision coming from finite size effects.

When $U$ increases, the groundstate changes into a more complicated many-body state with $n(\bm{k})$ having a non-zero contribution at the $\bm{\Gamma}$ point. This transition is accompanied by the change of the location of the minimum in momentum space.
In this regime the perturbation due to interactions can not be interpreted as small in terms of other parameters of our problem. The effect of interactions in this non-perturbative regime will not be considered in this work.

In simulations we do not observe the displacement of the condensation point due to interactions. This is explained by the fact that for weak interactions this effect is not noticeable enough to be observed because of the finite sizes of ED clusters, whereas in the regime of stronger interactions other effects occur earlier.

\section{Validity of the HFE convergence \label{sec:numFloquetCheck}}

We recall that all properties of the anisotropic Haldane model studied in Secs.~\ref{sec:effHaldaneModel_edgeModes} and \ref{sec:enginFmbec} correspond to the effective Hamiltonian, that we expand only up to the first order with contribution $\hat{H}_\text{eff}^{(1)}$. In this Section, we determine the conditions of validity for this approximation. More precisely, we want to compare the evolution of the system with the exact time-dependent Hamiltonian $\hat{H}(t)$ of Eq.~\eqref{eq:hamiltonian_timeDep_1p} to the evolution with the effective Hamiltonian of the anisotropic Haldane model.

\subsection{Second order terms in the HFE}

In order to study in details the convergence of the HFE, one needs to evaluate higher order terms in this perturbative expansion. In our case, i.e. for the time-dependent perturbation $\hat{V}(t)$ with only two non-vanishing Fourier components $\hat{V}^{(\pm 1)}$, the second order term is determined using the following identity:
\begin{equation}
\hat{H}^{(2)}_\text{eff} =
	\frac{1}{2\omega^2}
	\left(
	    \left[ \left[ \hat{V}^{(1)}, \hat{H}_0 \right], 
	    \hat{V}^{(-1)} \right]  + \text{h.c.} 
	\right)
\end{equation}
We study separately contributions coming from the term of frequencies (chemical potentials) $\omega_A$ and $\omega_B$ and the term of NN hoppings $t_1$, that are present in $\hat{H}_0$, and we denote them respectively by $\hat{H}^{(2)}_\text{eff} \left( \omega_B - \omega_A \right)$ and $\hat{H}^{(2)}_\text{eff}\left( t_1 \right)$.

If we consider the choice of phases depicted in Fig.~\ref{fig:anisHaldane_schemes},
the only effect of $\hat{H}^{(2)}_\text{eff} \left( \omega_B - \omega_A \right)$ 
consists in an asymmetric renormalization of chemical potentials, resulting in the generation of an effective Semenoff mass term. This effect vanishes if $\omega \gg \omega_A - \omega_B$.
$\hat{H}^{(2)}_\text{eff}\left( t_1 \right)$ generates NN and NNNN hoppings with complex amplitudes, that can be used in particular to obtain Chern insulators with Chern number greater than $1$. It also becomes irrelevant when $\omega \gg t_1$.

Moreover, if we consider the case of the many-body system with interactions weak enough, such that the Floquet approximation and the HFE are still valid, the interaction-dependent term  $\hat{H}^{(2)}_\text{eff}\left( U \right)$ will also appear at the second order in the perturbation theory. This term will lead to the generation of density-mediated NNN hoppings and $2^\text{nd}$ order hoppings such that 2 particles move to or from one particular lattice site at the same time. This term becomes negligible when $\omega \gg U$.
\begin{widetext}
\begin{align}
&\hat{H}^{(2)}_\text{eff} \left( \omega_B - \omega_A \right) =
\frac{V^2}{2\omega} \left( \frac{\omega_B - \omega_A}{\omega} \right)
\left[ \sum\limits_{\Braket{\Braket{ik}}} \cos(\Theta_{ik}) \left(
		\hat{a}^\dag_i \hat{a}_k + \hat{a}^\dag_k \hat{a}_i  
	\right) -
\sum\limits_{\Braket{\Braket{jl}}} \cos(\Theta_{jl}) \left(
		\hat{b}^\dag_j \hat{b}_l + \hat{b}^\dag_l \hat{b}_j  
	\right) 
\right]
\notag \\
&\hat{H}^{(2)}_\text{eff}\left( t_1 \right) =
\frac{V^2}{2\omega} \left( \frac{t_1}{\omega} \right)
\left( 
	\text{ NN and NNNN hoppings with complex amplitudes} 
\right)
\notag \\
&\hat{H}^{(2)}_\text{eff}\left( U \right) =
\frac{V^2}{2\omega} \left( \frac{U}{\omega} \right)
\left( 
	\text{density-mediated NNN hoppings and } 2^\text{nd} \text{ order hoppings with complex amplitudes}
\right)
\end{align}
\end{widetext}

\subsection{Numerical convergence}

\begin{figure*}
\includegraphics[width=0.9\textwidth]
{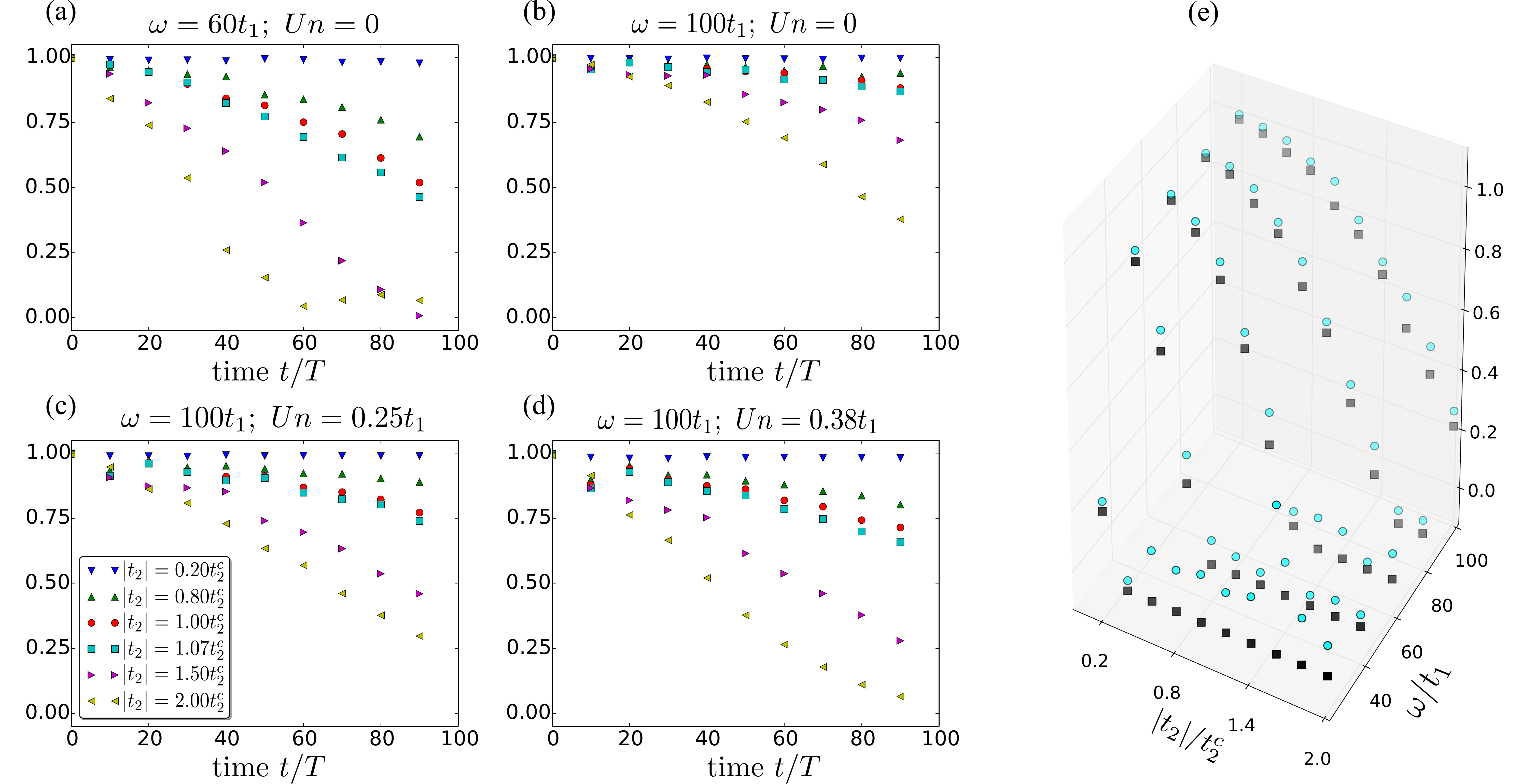}
\caption{Observable $F(t) = \left|\Braket{\Psi(t) | \Psi_\text{eff}(t)}\right|$ for the time evolution over $T = 100$ periods \textbf{(a-b)} Stroboscopic values of the fidelity in the single-particle case ($U=0$) for two different values of $\omega$. \textbf{(c-d)} Stroboscopic values of the fidelity in the many-body case for two different values of $U$, with $\omega = 100 t_1$. \textbf{(e)} Diagram showing minima (black squares) and local maxima (blues circles) of $F(t)$ for different values of $t_2$ and $\omega$ in the single-particle case ($U=0$). All simulations are performed with randomly distributed initial state $\Ket{\Psi_\text{rand}}$.}
\label{fig:numHfeConvergencePlots}
\end{figure*}

We estimate more quantitatively, which values of the modulation frequency $\omega$ are big enough, compared to $t_1$ and $U$, so that we can neglect all second order contributions in the HFE. For clarity, we consider the case of $\omega_A = \omega_B$, such that the term $\hat{H}^{(2)}_\text{eff} \left( \omega_B - \omega_A \right)$ is exactly zero. Therefore, we perform numerical simulations of the exact time-evolution of the model. We initially prepare the system in the state $\Ket{\Psi(t=0)} = \Ket{\Psi_0}$ (in the following we will consider three different choices for this initial state) and we use exact diagonalization technique to calculate states $\Ket{\Psi(t)}$ and $\Ket{\Psi_\text{eff}(t)}$ at time $t$, evolved with $\hat{H}(t)$ and $\hat{H}_\text{eff}$ respectively, where the last Hamiltonian is calculated up to the first order in the HFE (and thus corresponds to the Hamiltonian of the anisotropic Haldane model). As far as two states propagate in time, they deviate one from another. This deviation is captured by the "fidelity" $F(t)$ defined as follows:
\begin{equation}
F(t) = \left|\Braket{\Psi(t) | \Psi_\text{eff}(t)}\right|\;.
\end{equation}
From the results of Sec.~\ref{sec:enginFmbec}, we notice that we are also interested in the regime where the effectively generated NNN hopping amplitude $t_2 = \frac{V^2}{2\omega}$ becomes of the order of $t_1$, which implies that means that the amplitude $V$ behaves as $\sqrt{\omega t_1}$ in the limit $\omega \rightarrow \infty$. Thus, in order to have a more complete description, we perform simulations for different values of $t_2$. Once $t_2$ is fixed, we deduce the value of the Floquet modulation amplitude using $V = \sqrt{2 \omega t_2}$. In particular, the critical value $V_c$ required to reach the \textit{ZM} -- \textit{FM} phase transition at fixed frequency $\omega = 60 t_1$ is $V_c \approx 8.57 t_1$.

\begin{figure}
\includegraphics[width=0.35\textwidth]
{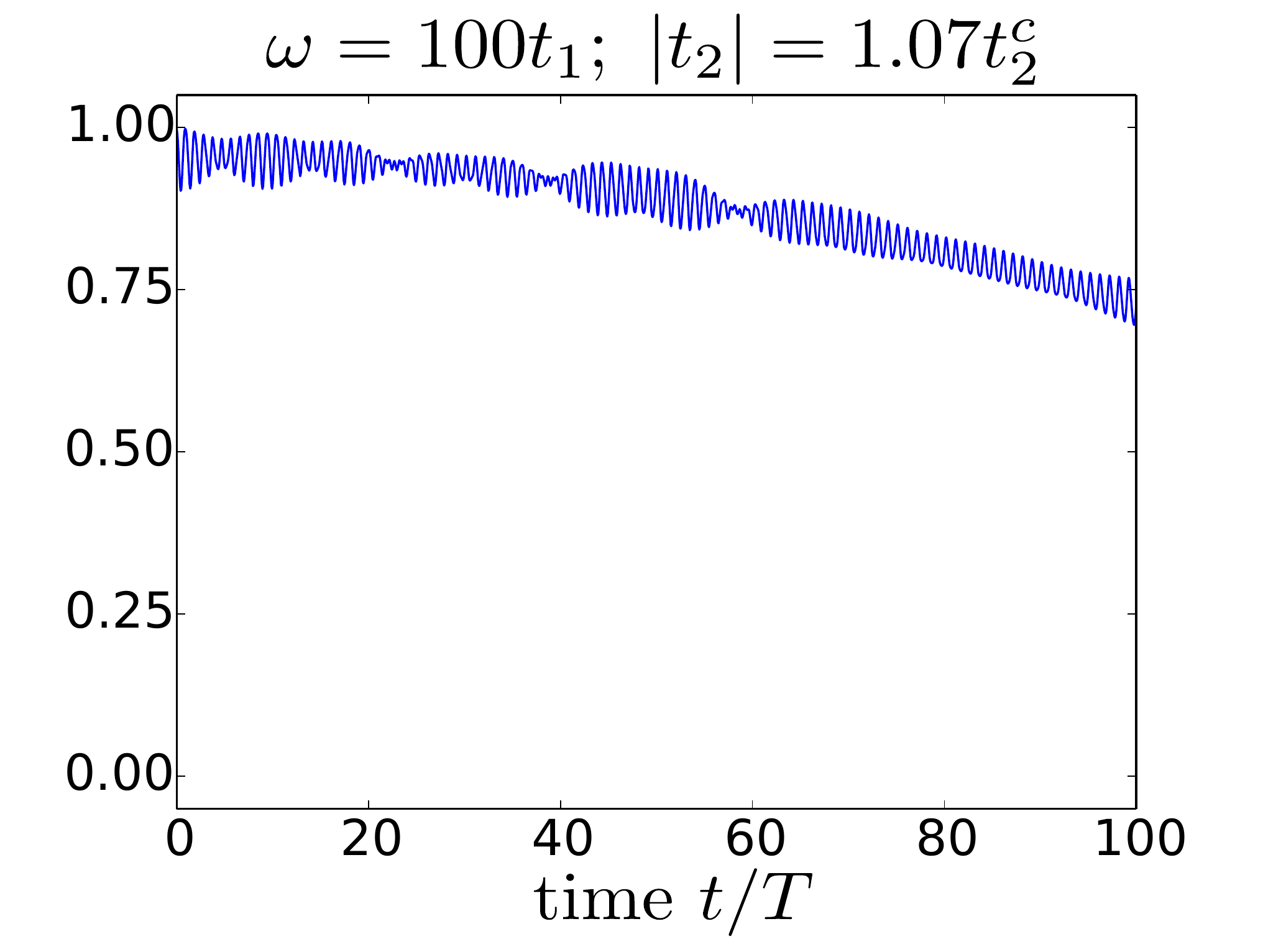}%
\caption{Numerical measurement of the fidelity $F(t) = \left|\Braket{\Psi(t) | \Psi_\text{eff}(t)}\right|$ for the time evolution over 100 periods in the single-particle case ($U=0$). Exact time-dependent evolution is performed for $V = \sqrt{2\omega t_2}$.}
\label{fig:fidelityTypShape}
\end{figure}

To start with, we show the calculations performed in the single-particle case. Fig.~\ref{fig:fidelityTypShape} represents the typical shape of the fidelity for the time evolution over 100 periods. The first important observation is related to the fast oscillation of $F(t)$, forming an "envelop" for its propagation in time. These oscillations are due to intra-periodic submotions of the system. They are not captured by $\hat{H}_\text{eff}$, but are described by Kick operators $\hat{K}_\text{eff}$ (see Eq.~\eqref{eq:effectiveHamiltonian}).

Secondly, during the time evolution over an integer number of periods, the value of intra-periodic maxima of $F(t)$ decreases. This effect originates from higher order terms in the HFE. Errors due to neglected terms are accumulated during each period, leading to an exponential decrease of $F(t)$ with time. The total error becomes smaller when we increase the ratio $\omega/t_1$ or decrease $t_2/t_1$. This effect is observed more quantitatively in Fig.~\ref{fig:numHfeConvergencePlots}(a-b), where we plot the fidelity for different values of parameters $\omega$ and $t_2$. For clarity, we do not show the entire time-evolution, but only the value of $F(t)$ at few stroboscopic times. In Fig.~\ref{fig:numHfeConvergencePlots}(e), we also plot the diagram of minima and local intra-periodic maxima of the fidelity, corresponding to the lowest position of the "envelop" of $F(t)$.

Another important point, not captured by Floquet theory and the HFE, and which appears in simulations, is the fact that the time evolution with both effective and exact Hamiltonians depends on the initial state $\Ket{\Psi_0}$. In our numerical simulations we consider three different types of initial states: the groundstate of the effective Hamiltonian $\Ket{GS_\text{eff}}$, the groundstate of the unperturbed Hamiltonian $\Ket{GS_{0}}$ and a state $\Ket{\Psi_\text{rand}}$ with a random wavefunction.

All properties related to the fidelity, described above, are observed in the case of the state $\Ket{\Psi_\text{rand}}$. If however we consider the state $\Ket{GS_{0}}$, we do not observe any noticeable time-evolution. In the case of the state $\Ket{GS_\text{eff}}$ in the regime when the minimum of the lowest band is not localized at point $\bm{\Gamma}$, we clearly observe intra-periodic sub-motions of the system. However, the envelop of $F(t)$ is located close to the maximum and does not decay with time. This can be interpreted in terms of the symmetry of these states with respect to Hamiltonians. For example, if the initial state is one of two eigenstates at momentum $\bm{k} = 0$, it will always stay an eigenstate of both effective, exact time-dependent and unperturbed Hamiltonians. Thus, we will not observe any evolution of the fidelity.

In order to prepare the basis for future investigations, we perform similar numerical tests for fidelities of the HFE in the many-body case. We are particularly interested in how the convergence of the HFE is modified by Bose-Hubbard interactions. As we already mentioned, interactions lead to highly non-trivial effects such as heating or decoherence. In Fig.~\ref{fig:numHfeConvergencePlots}(c-d) we plot $F(t)$ for two different values of $Un$. We see that the effect of $U$ leads to the faster decay of the fidelity.

\section{Ladder Geometry \label{sec:ladders}}

\begin{figure}
\includegraphics[width=0.40\textwidth]
{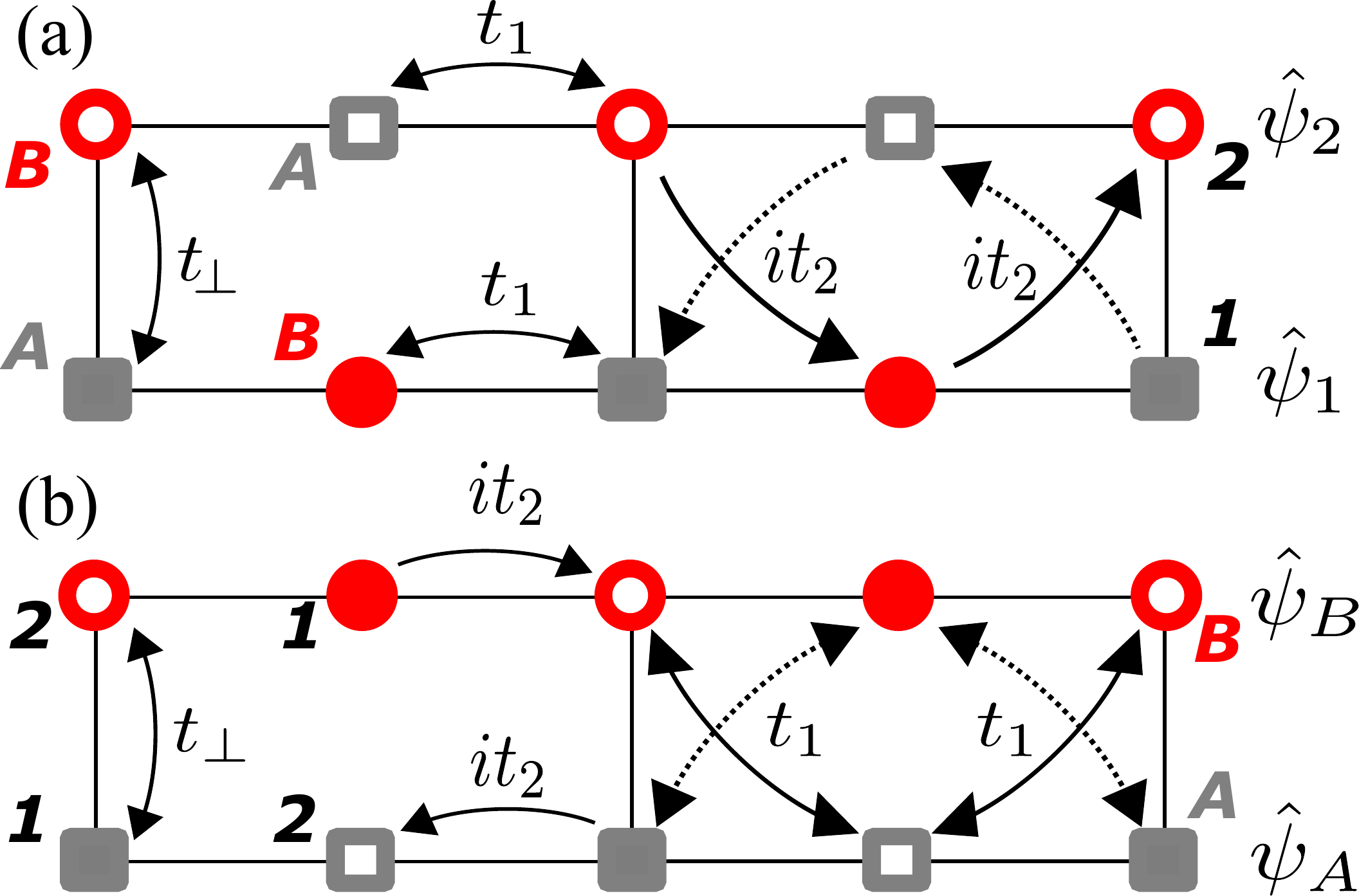}
\caption{Strip (ladder) geometry consisting of two coupled chains. Each site of the unit cell of this ladder is characterized by the chain index $1$ or $2$ (represented in the picture by the fact that the figure is either filled or not) and the sublattice index $A$ or $B$ (represented by the color and the shape of the figure). In the continuum limit, depending on the ratio $t_1/t_2$ we would prefer either to describe the system in terms of \textbf{(a)} operators $\hat{\psi}_1$ and $\hat{\psi}_2$ if $t_1 \gg t_2$, or \textbf{(b)} operators $\hat{\psi}_A$ and $\hat{\psi}_B$ if $t_1 \ll t_2$.}
\label{fig:strip_geom}
\end{figure}

Many theoretical and experimental works were performed during last decades using the ladder geometry, allowing for a strip version of two-dimensional lattices and the efficient theoretical tools of quasi one-dimensional systems such as bosonization and numerical techniques. In particular, in bosonic systems, questions related to the Mott insulator -- superfluid phase transition, effects induced by the magnetic field, such as Meissner effect and apparition of one dimensional equivalent of a vortex lattice \cite{Orignac2001MeissnerMottLadders, Dhar2012MeissnerMottLadders, Petrescu2013MeissnerMottLadders, Tokuno2014MeissnerMottLadders, Piraud2015MeissnerMottLadders, Greschner2015MeissnerMottLadders, Orignac2016MeissnerMottLadders} or also Laughlin bosonic phases \cite{Petrescu2014LauglinBosonicLadders} were addressed.
Experimental realization of bosonic ladders, giving opportunity to study the rich and profound physics of these systems were done recently, using laser assisted tunneling~\cite{Atala2014bosonicLaddersUltracoldAtoms} to create artificial gauge fields. 
For fermionic systems, the observation of intriguing chiral edge states in systems with synthetic dimensions was reported in Ref.~\onlinecite{Mancini2015FermionicChiralEdgeStates}, in relation to recent theoretical works \cite{Lacki2016qheSyntDimms, Barbarino2016qheSyntDimms, Taddia2016topFracPumpArtificialDimms}.

Motivated by this context, we consider in this section a ladder version of the anisotropic Haldane model introduced in the preceding sections. We reformulate the problem of Sec.~\ref{sec:enginFmbec} in the ladder geometry of Fig.~\ref{fig:strip_geom}. We consider the distance between NN being equal to $a$. The unit cell of such ladders is formed by 4 distinct sites, that we will distinguish by chain index $\nu \in \{0, 1 \}$ and by sublattice index $c \in \{A, B \}$, with the convention that $\hat{c}_{\nu,i}$ is either $\hat{a}_{\nu,i}$ or $\hat{b}_{\nu,i}$, reminiscent of the 2D formulation of the problem. The effective Hamiltonian evaluated up to the first order in the HFE can be conveniently written in terms of 4 distinct terms
$ \hat{H} = \sum\limits_{\nu, c}
	\hat{H}^{U}_{\nu, c} +
\sum\limits_{\nu}
	\hat{H}^{\parallel}_\nu + 
\sum\limits_{c} 
	\hat{H}^{t_2}_{c} +
\hat{H}^{\perp} $ 
expressed as follows:
\begin{widetext}
\begin{align}
\label{eq:stripGeomHamiltonians}
\hat{H}^{\parallel}_1 + \hat{H}^{\parallel}_2 &=
	- t_1 \sum\limits_{i} \left( 
	\hat{a}^\dag_{1,2i} \hat{b}_{1,2i+1} +
	\hat{a}^\dag_{1,2i} \hat{b}_{1,2i-1} + 
	\hat{b}^\dag_{2,2i} \hat{a}_{2,2i+1} +
	\hat{b}^\dag_{2,2i} \hat{a}_{2,2i-1} +
	\text{h.c.} \right) 
\notag \\
\hat{H}^{t_2}_{A} + \hat{H}^{t_2}_{B} &=
	- i t_2
	\sum\limits_{i}
	\left( \hat{a}^\dag_{1,2i} \hat{a}_{2,2i+1} + 
	\hat{a}^\dag_{2,2i-1} \hat{a}_{1,2i} +
	\hat{b}^\dag_{2,2i} \hat{b}_{1,2i-1} + 
	\hat{b}^\dag_{1,2i+1} \hat{b}_{2,2i} - \text{h.c.} 
	\right)
\notag \\
\hat{H}^{\perp} &= - t_{\perp} \sum\limits_{i} 
	\left( \hat{a}^\dag_{1,2i} \hat{b}_{2,2i} + \hat{b}^\dag_{2,2i} \hat{a}_{1,2i}  \right)
\notag \\
\hat{H}^{U}_{\nu, c} &=	
	\frac{U}{2}\sum\limits_{i} \hat{n}_{\nu,c,i} (\hat{n}_{\nu,c,i}-1)
\end{align}\;
\end{widetext}
in which the spatial index $i$ runs over $\left[ 1, 2, \dots, L/2 \right]$, with $L$ the length of the ladder containing $L/2$ unit cells. If we compare the strip geometry to the 2D geometry, we should consider $t_{\perp} = t_1$. Yet, we will study a more general case with arbitrary hopping $t_{\perp}$.

\subsection{Single-particle spectrum basis}

\begin{figure}
\includegraphics[width=0.48\textwidth]
{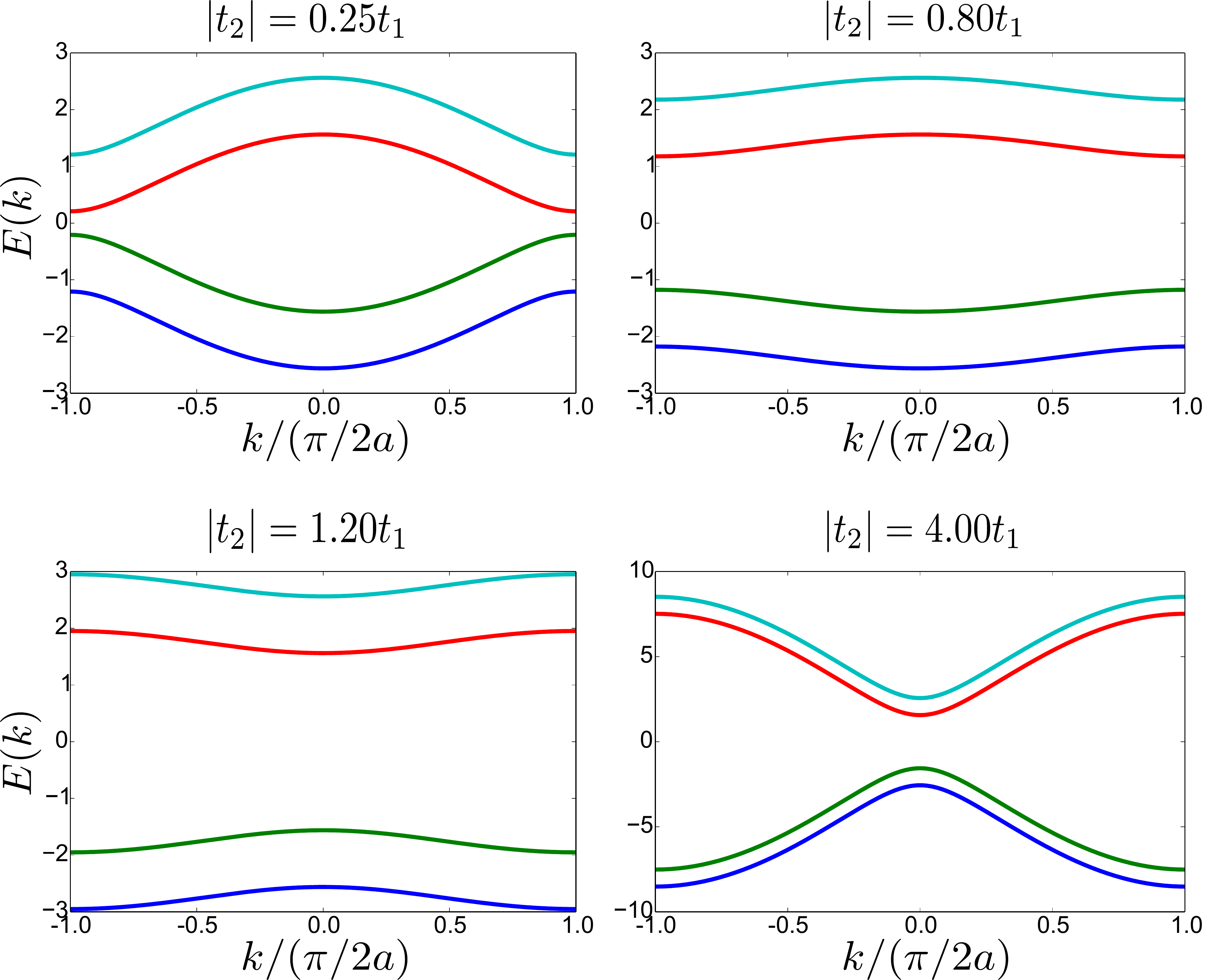}
\caption{Single particle spectra of the ladder model ($U=0$). All figures have parameters $t_1 = t_{\perp} = 1$. We see that as $t_2$ increases, the minima of the bands moves from $k = 0$ to $k = \pm \frac{\pi}{2a}$. The gap between middle bands closes at $t_2 = 0$.}
\label{fig:strip_1prt}                                                                                                                                                                          
\end{figure}

We first calculate the band structure $\epsilon_0(k)$ of the non-interacting system by going to the momentum space. The Hamiltonian is rewritten in terms of $4\times4$ matrices $\mathcal{H}(k)$ as follows:
$\hat{H} = -\sum\limits_{k}
\hat{\psi}^\dag_{k} \cdot \mathcal{H}(k) \cdot \hat{\psi}_{k}
$, where 
$k \in \left[-\frac{\pi}{2a}, \frac{\pi}{2a} \right]$ is the vector in the 1D BZ,
$\hat{\psi}_{k} =
\left(
	\hat{a}_{1, k}, \hat{b}_{1, k},
	\hat{a}_{2, k}, \hat{b}_{2, k}
\right)^t$
and
\begin{equation}
\label{eq:Hk_strip_1p}
\hspace*{-7pt}
\hat{\mathcal{H}}(k) =
\begin{pmatrix}
	0 & 2t_1 \cos(a k) & -2t_2 \sin(a k) & t_{\perp} \\
	2t_1 \cos(a k) & 0 &	0 & 2t_2 \sin(a k) \\
	-2t_2 \sin(a k) & 0 & 0 & 2t_1 \cos(a k) \\
	t_{\perp} & 2t_2 \sin(a k) & 2t_1 \cos(a k) & 0 \\
\end{pmatrix}
\end{equation}
In Fig.~\ref{fig:strip_1prt}, we show the spectrum $\epsilon_0(k)$ for various $t_2$. If we consider for instance $t_2 = 0$, we find that the spectrum 
$ \epsilon_0(k) = \pm \frac{1}{2} \left[ t_{\perp} \pm \sqrt{t_{\perp}^2 + 16 t_1^2 \cos^2(a k)} \right] $, 
to be compared to the standard ladder bands $\epsilon_0(k) = \pm t_{\perp}-2 t_{1} \cos\left( a k \right)$. 
In the general case, the four eigenvalues of $\mathcal{H}(k)$ are 
\begin{equation}
\epsilon_0(k) = \pm \frac{1}{2} \left[ t_{\perp} \pm \sqrt{t_{\perp}^2 + 16 t_1^2 \cos^2(a k) + 16 t_2^2 \sin^2(a k) } \right]
\end{equation}
From this equation and Fig.~\ref{fig:strip_1prt}, we observe that there is a transition at $t_1=t_2$, separating the situation $t_2 < t_1$, in which the minimum of the lowest band is located at $k=0$, from the situation $t_2 > t_1$, where the minimum of the lowest band is $k = \pi/2a$ (up to a reciprocal lattice vector).
This is thus analogue to the ZM--FM transition of the 2D version of the model, except that the minimum at $k = \pi/2a$ does not move when changing $t_2$.
This transition is actually related to the duality in the model, appearing in Fig~\ref{fig:strip_geom}, in which $t_1 \leftrightarrow it_2$ and one site each two is exchanged. This duality maps local kinetic energy onto current operators and explains the $t_1=t_2$ transition point. 

It is lastly important to notice that, at this very transition point $t_1 = t_2$, the model displays flat bands since the four energies become $ \epsilon_0(k) = \pm \frac{1}{2} \big[ t_{\perp} \pm \sqrt{t_{\perp}^2 + 16 t_1^2 } \big] $ and are thus independent of $k$. Such peculiar band structure makes it difficult to analyze the effect of interactions on the model and we leave this question for possible future investigations.
Consequently, as bosonization requires to linearize the bands and numerics become challenging for flat band models, we discuss in what follows the regimes $t_2 \ll t_1$ and $t_2 \gg t_1$, expecting that a third phase would appear around the transition point. This gives an account on how the two phases emerge in the presence of interactions.

\subsection{Low energy continuum description and two competing phases}

In order to describe the behavior of the interacting system in \textit{ZM} and \textit{FM} phases, far from the regime of flat bands, we write Hamiltonian~\eqref{eq:stripGeomHamiltonians} in the continuum limit. If one considers only one chain for instance, its low-energy description falls into the universality class of Tomonaga-Luttinger liquids. Excitations are collective sound modes with linear dispersion and are described using the "harmonic fluid approach" also known as "bosonization" \cite{Haldane1981a, Gogolin2004bosonization, Giamarchi2003QpIn1D, Cazalilla2011OneDimBosons}. 
In the geometry under study and setting $x = ja$ with $a$ the lattice spacing, one has the freedom to define either the set of operators
$ \hat{\psi}_{1(2)} (x) = \hat{c}_{1(2),j}/\sqrt{a} $,
corresponding to the lower/upper chain prescription, or the operators
$ \hat{\psi}_{A(B)} (x) = \hat{a}(\hat{b})_{j}/\sqrt{a} $ that correspond to the sublattice prescription.
This choice of prescription is schematically represented in Fig.~\ref{fig:strip_geom}. Bosonic creation operators $ \hat{\psi}^\dag_{\nu} (x) $ are then written in terms of new bosonic fields $\hat{\theta}_{\nu}(x)$ and $\hat{\phi}_{\nu}(x)$ via the following relation:
\begin{equation}
\label{eq:bosonisationForBosons}
\hat{\psi}^\dag_{\nu} (x) = 
\left( \rho_0 - \frac{1}{\pi} \nabla \hat{\phi}_\nu(x) \right)^{1/2} 
\sum\limits_{p}
	e^{i2p \left( \pi \rho_0 x - \hat{\phi}_\nu(x) \right)}
	e^{-i\hat{\theta}_\nu(x)}\;.
\end{equation}
The $\hat{\theta}_{\nu}(x)$ are phase fields and the $\hat{\phi}_{\nu}(x)$ are the long wavelength density excitation, such that
$ - \nabla \hat{\phi}_{\nu}(x)/\pi = \hat{\rho}_{\nu}(x)- \rho_0 $, in which 
$\hat{\rho}_{\nu}(x)$ is the density operator and $\rho_0$ its mean value in the ground-state. 
If the system is translationally invariant, $\rho_0 = n/a$ with $n=N/N_s$ the filling. 
Operators $\hat{\theta}_{\nu}(x)$ and $\hat{\phi}_{\nu}(x)$ satisfy the following commutation relation:
\begin{equation}
\left[ \hat{\phi}_{\mu}(x), \hat{\theta}_{\nu}(x') \right] =
i \frac{\pi}{2} \delta_{\mu \nu} \operatorname{Sign}(x-x')\;.
\end{equation}
The oscillating contribution in Eq.~\eqref{eq:bosonisationForBosons} reflects the ordering in the lattice description of the model (particles tend to develop a crystal-like structure). If one averages the density over the distances large compared to $a$, only the $p=0$ term will remain. This is the simplification that we will consider in the following.

\subsection{Strong $t_1$ phase}

When NNN hopping amplitude $t_2$ is small compared to $t_1$, the lower/upper chain prescription is well suited. 
This allows us to express
$ \hat{H}^{L}_{\nu} =
\hat{H}^{\parallel}_{\nu} + \hat{H}^{U}_{\nu, A} + \hat{H}^{U}_{\nu, B} $ as follows:
\begin{equation}
\hat{H}^{L}_{\nu} = \int 
\frac{dx}{2 \pi} 
\left(
v K \big| \nabla \hat{\theta}_{\nu}(x) \big|^2 + 
\frac{v}{K} \big| \nabla \hat{\phi}_{\nu}(x) \big|^2
\right)
\end{equation}
where the speed of sound is $v$ and the Luttinger parameter is $K$. 
In the weakly interacting limit, they are identified as $v = \rho_0 \sqrt{t_1 U}$ and $K = \sqrt{t_1/U}$. 
Furthermore, the Luttinger parameter satisfies $K > 1$ for finite repulsive short-range interactions and $K = 1$ in the hard core limit. 
The two chains are coupled by the following terms
\begin{equation}
\hat{H}^{\perp} = - \rho_0 t_{\perp}\int dx
\cos \left[ \hat{\theta}_{1}(x) - \hat{\theta}_{2}(x) \right]\;,
\end{equation}
and
\begin{align}
\hat{H}^{t_2}_A + \hat{H}^{t_2}_B = \rho_0 t_2
 \int dx \bigg( 
	 &\sin \left[ \hat{\theta}_{2}(x+a) - \hat{\theta}_{1}(x) \right] \notag\\
	+& \sin \left[ \hat{\theta}_{1}(x) - \hat{\theta}_{2}(x-a) \right]\notag\\
	+&\sin \left[ \hat{\theta}_{2}(x) - \hat{\theta}_{1}(x+a) \right]\notag\\
	+& \sin \left[ \hat{\theta}_{1}(x-a) - \hat{\theta}_{2}(x) \right]
\bigg) \;.
\end{align}
If $t_{\perp}$ is strong (compared to $t_1$), phases $\hat{\theta}_\mu(x)$ will be pinned by the term $\hat{H}^{\perp}$ in such a way that $\braket{\hat{\theta}_{1}(x)} = \braket{\hat{\theta}_{2}(x)} = \text{const}$, where $\Braket{.}$ is taken in the ground-state. Such configuration also implies that $\braket{\hat{H}^{t_2}_c} = 0$ and bosons on two chains form a quasi-condensate at the point $k = 0$ in the BZ. 
The expectation value of $\hat{\theta}_{\nu}(x+a) - \hat{\theta}_{\nu}(x)$ and 
$\hat{\theta}_{2}(x+a) - \hat{\theta}_{1}(x)$ is related to the value of densities of local currents between respectively NN and NNN sites. 
One can deduce in particular that to the lowest order in $a$ all currents that flow in the right direction (with increasing $x$) can be expressed as
\begin{align}
J^{\parallel, 1}_{AB, r} &= - J^{\parallel, 2}_{AB, r} =
- 2 \rho_0 \operatorname{Im} \left[ 
t_1 \Braket{e^{i \left( \hat{\theta}_{1}(x) - \hat{\theta}_{1}(x+a) \right)}} 
\right] = 0
\notag \\
J_{AA, r} &= - J_{BB, r} = 
- 2 \rho_0  \operatorname{Im} \left[ 
it_2 \Braket{e^{i \left( \hat{\theta}_{1}(x) - \hat{\theta}_{2}(x+a) \right)}}
\right] \notag \\ 
& = - 2 \rho_0 t_2
\end{align}
Indices $AB$ and $AA$ correspond respectively to NN and NNN currents, $1$ and $2$ are chain indices and $r$ refers to the right direction. Interchain currents are zero everywhere:
\begin{equation}
J^{\perp}_{AB} = - 2 \rho_0  \operatorname{Im} \left[ 
t_1 \Braket{e^{i \left( \hat{\theta}_{2}(x) - \hat{\theta}_{1}(x) \right)}} 
\right] = 0,
\end{equation}
which is a property related to the Meissner phase in bosonic ladders, observed in particular in the experimental realization reported in Ref.~\onlinecite{Atala2014bosonicLaddersUltracoldAtoms}.

\begin{figure}
\includegraphics[width=0.48\textwidth]
{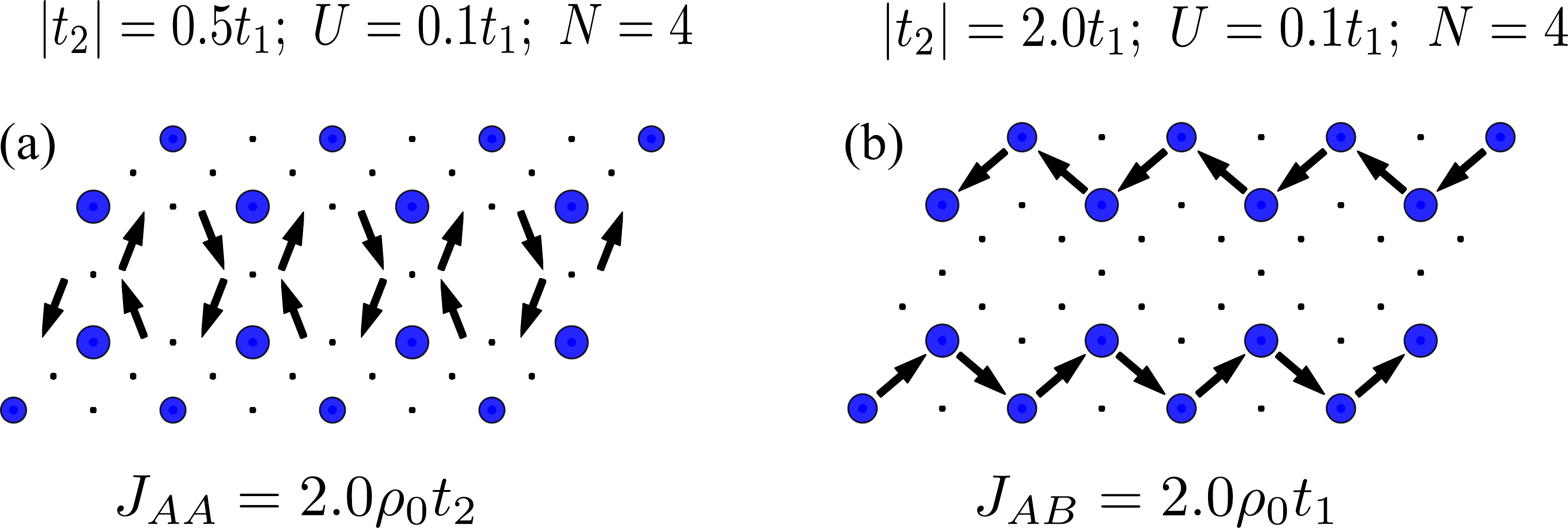}
\caption{Numerical simulations showing the LDOS $\rho_i$ (via scaling of blue circles) and the lattice current density $J_{ij}$ (black arrows) in the ground-state of the system in ladder geometry with $8\times2$ sites. \textbf{(a)} The phase $t_1 > t_2$ is characterised by $|J^{\parallel}_{AB}| = 0$ and $|J_{AA}| \approx 2 \rho_0 t_2$. \textbf{(b)} The phase $t_1 < t_2$ is characterised by $|J^{\parallel}_{AB}| \approx 2 \rho_0 t_1$ and $|J_{AA}| = 0$. All simulations were performed in the system with PBC along the $x$ axis.}
\label{fig:strip_currents}                                                                                                                                                                          
\end{figure}

We illustrate this scenario by ED calculations.
The numerical results are shown in Fig.~\ref{fig:strip_currents}(a) and display the expected current pattern for $t_2=t_1/2$.

\subsection{Strong $t_2$ phase}

From the other side, when $t_2$ is strong, terms $\hat{H}^{t_2}_A$ and $\hat{H}^{t_2}_B$ favor the generation of two decoupled intercrossing chains characterized by
$ \braket{\hat{\theta}_{2}(x+a) - \hat{\theta}_{1}(x)} = 
\braket{\hat{\theta}_{1}(x) - \hat{\theta}_{2}(x-a)} = -\pi/2 $ 
(i.e. the phase decreases with $x$), and
$ \braket{\hat{\theta}_{1}(x+a) - \hat{\theta}_{2}(x)} = 
\braket{\hat{\theta}_{2}(x) - \hat{\theta}_{1}(x-a)} = \pi/2 $
(i.e. the phase increases with $x$) for all $x = 2ja$.
In order to simply describe this phase, one uses the sublattice prescription, considers operators $\hat{\theta}_{A/B}(x)$ and $\hat{\phi}_{A/B}(x)$
and performs the gauge transformation:
$ \hat{\theta}_{A/B}(x) \rightarrow \hat{\theta}'_{A/B}(x) = 
\hat{\theta}_{A/B}(x) \pm \pi x/2 a $. This leads to
\begin{equation}
\hat{H}^{t_2}_c + \hat{H}^{U}_{1,c} + \hat{H}^{U}_{2,c} = 
 \int \frac{dx}{2 \pi}
\bigg(
u K \big| \nabla \hat{\theta}'_{c}(x) \big|^2 + 
\frac{u}{K} \big| \nabla \hat{\phi}_{c}(x) \big|^2
\bigg)\;,
\end{equation}
with $u,K$ the corresponding parameters.
The two chains are coupled by $t_1$ and $t_{\perp}$ terms in the following way:
\begin{equation}
\hat{H}^{\perp} = - t_{\perp} \rho_0 \int dx 
\cos \left[ \hat{\theta}_{A}(x) - \hat{\theta}_{B}(x) \right]\;,
\end{equation}
and
\begin{align}
\hat{H}^{\parallel}_1 + \hat{H}^{\parallel}_2 = -t_1 \rho_0
\notag \int dx \bigg( 
	 &\cos \left[ \hat{\theta}_{B}(x+a) - \hat{\theta}_{A}(x) \right] \\
	+ &\cos \left[ \hat{\theta}_{A}(x) - \hat{\theta}_{B}(x-a) \right]\notag \\	
	+&\cos \left[ \hat{\theta}_{B}(x) - \hat{\theta}_{A}(x-a) \right] \notag \\
	+ &\cos \left[ \hat{\theta}_{A}(x+a) - \hat{\theta}_{B}(x) \right]
\bigg) 
\end{align}
The ground-state in the strong $t_2$ limit corresponds to the quasi-condensation of bosons at the point $k = \pm \frac{\pi}{2a}$ (defined up to a reciprocal lattice vector $\pi/a$) in the BZ. The effect of the $t_{\perp}$ coupling consists in pinning the phases of the two chains. One remarks that, according to the duality, $t_1$ plays the same role as $t_2$ in the strong $t_1$ phase.
We are also interested in how expectation values of the local current are modified. Straightforward calculations imply that in the strong $t_2$ phase currents (expressed in the same way as previously) become 
\begin{align}
J^{\parallel, 1}_{AB, r} &= - J^{\parallel, 2}_{AB, r} =
- 2 \rho_0 \operatorname{Im} \left[ 
t_1 \Braket{e^{i \left( \hat{\theta}_{1}(x) - \hat{\theta}_{1}(x+a) \right)}} 
\right] = 2 \rho_0 t_1
\notag \\
J_{AA, r} &= - J_{BB, r} = 
- 2 \rho_0  \operatorname{Im} \left[ 
it_2 \Braket{e^{i \left( \hat{\theta}_{1}(x) - \hat{\theta}_{2}(x+a) \right)}}
\right] = 0
\notag \\
J^{\perp}_{AB} &= - 2 \rho_0  \operatorname{Im} \left[ 
t_1 \Braket{e^{i \left( \hat{\theta}_{2}(x) - \hat{\theta}_{1}(x) \right)}} 
\right] = 0
\end{align}
These analytical predictions are also illustrated by ED simulations gathered in Fig.~\ref{fig:strip_currents}(b).

\subsection{Related model}

\begin{figure}
\includegraphics[width=.35\textwidth]
{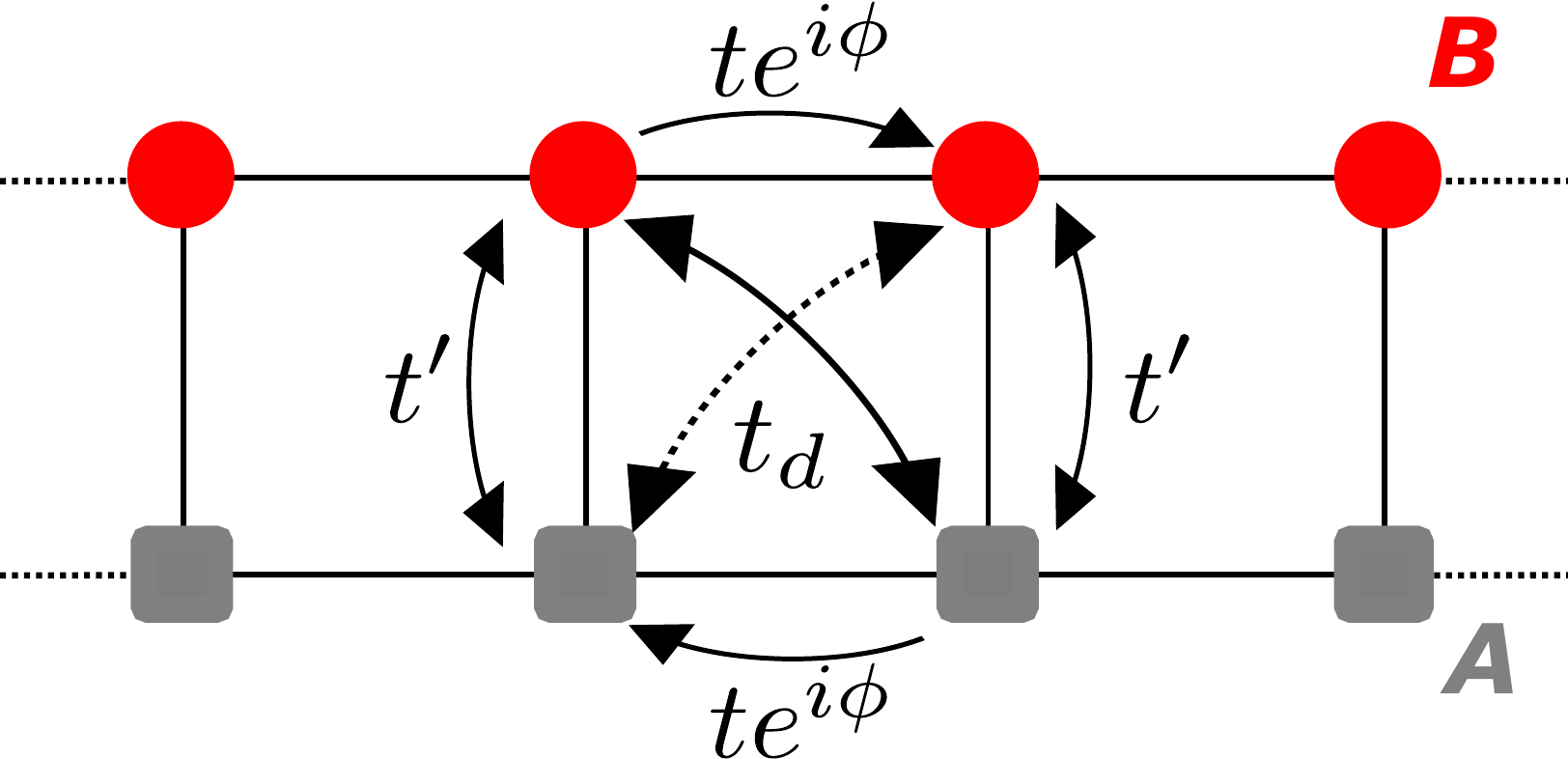}
\caption{Schematic representation of the ladder model of Refs.~\onlinecite{Hugel2014chiralLadders, Sticlet2014CreutzModel}, which can be generated by using the Floquet theory and the HFE, starting from the Hamiltonian \eqref{eq:hamiltonian_timeDep_1p}.}
\label{fig:stripGeom_Hugel}
\end{figure}

We emphasize that many other effective models with non-trivial profound physics could be generated using the time-dependent model described by Eq.~\eqref{eq:hamiltonian_timeDep_1p}. 
In particular, one could consider the square lattice with phases $\theta_{ij}$ distributed in the following way: $\theta_{ij} = \pi/2$ for all horizontal hoppings from the right to the left and $\theta_{ij} = 0$ for all vertical hoppings from the bottom to the top. The resulting model is geometrically equivalent to the model of Refs.~\onlinecite{Hugel2014chiralLadders, Sticlet2014CreutzModel} (in the last Reference it was also referred as the Creutz model) and sketched in Fig.~\ref{fig:stripGeom_Hugel}.
This models also possesses a one to two quantum wells transition at the bosonic ground-state level and non-trivial topological properties related to the generation of an effective spin-orbit coupling (due to $t_d$ term that couples different chains of the ladder at neighbouring positions $j$).

\section{Conclusion}

To summarize, we have proposed a way to engineer photonic Haldane type Chern insulators on the graphene lattice simulating artificial gauge fields, based on the Floquet theory and a related high-frequency expansion. We have shown how the topology of Bloch bands subsists to anisotropic bonds coupling NNNs (in the present case, closed loops of vector potentials involve kite geometries rather than triangles), but the critical point of the topological phase transition is modified. We have tested the validity of the Floquet theory and the high-frequency expansion based on a time-dependent exact diagonalization scheme on small clusters. These results are in direct relation with recent experiments in ultra-cold atoms \citep{Jotzu2014haldaneShaking} and photon and circuit systems \cite{Roushan2016supercondTriang}. Focusing on BEC configurations in optical lattices and a Gross-Pitaevski approach, then we have shown how using the s-band of optical lattices (instead of the p-band) realizes a BEC analogue of FFLO states predicted for fermionic systems in the Bardeen-Cooper-Schrieffer limit. At a more general level, we have shown how the quantum phase transition between the zero-momentum BEC (ZM BEC) phase and the finite-momentum BEC (FM BEC) phase can be observed using a time-dependent protocol periodic in time. This is an example of Floquet quantum phase transitions. The transition can be detected via an indirect measurement of Meissner currents along NNNs, averaged over one period of Floquet perturbation, or by focusing on the momentum distribution of bosons. We find that the nature of the quantum phase transition is different in the isotropic \cite{VassicPetrescu2015} and anisotropic contexts. In the latter case studied in the present paper, the transition is of second-order type, which is exemplified by the continuity of Chiral currents at the transition. Closed loops of Chiral currents also involve lozenges instead of triangles. We have shown how interactions affect the band structure and the momentum distribution across the quantum phase transition. The effect of quite large interactions on the Floquet theory leading potentially to heating and decoherence effects will be studied separately; these effects must be addressed before considering the possibility to realize Mott physics with Floquet schemes. The present engineered gauge for the vector potentials also allows tractable solutions in ladder (strip) geometries. We have shown, using bosonization and exact diagonalization, that a similar chiral bosonic quantum phase transition occurs in this case. However, the vicinity of the phase transition, which exhibits flat bands, is more complex and it will consequently be studied separately. Similar Meissner currents have already been reported in bosonic ladders \cite{Atala2014bosonicLaddersUltracoldAtoms} by analogy to superconductors. This time-dependent approach could also be extended to simulate spin-orbit models in relation with iridate materials \cite{KaneMele2005model, RachelLeHur2010topInsHubbard, Tianhan2013, Tianhan2015ffloSupercondKitaevHeisenberg}.

\subsubsection*{Acknowledgements}

This work has benefited from discussions and collaborations with Ivana Vasic, Alexandru Petrescu and Walter Hofstetter. We also acknowledge discussions with Immanuel Bloch, Beno\^{i}t Dou\c{c}ot, Tianhan Liu, Belen Paredes, Arun Paramekanti, Leticia Tarruell, Ronny Thomale, Julien Gabelli, J\'erome Est\`eve, Lo\"{i}c Herviou, Lo\"{i}c Henriet and Aleksey Fedorov. We acknowledge discussions at CIFAR meetings in Canada and at conferences in University of Cergy-Pontoise. We acknowledge support by the German Science Foundation (DFG) FOR2414. We also acknowledge support from the PALM Labex, Paris-Saclay, Grant No. ANR-10-LABX-0039.

\appendix

\section{Effective Haldane Hamiltonian and dominant terms in Dyson series for observables}
\label{app:DysonPerturbationTheory}

It is known that the Floquet Hamiltonian obtained through the Floquet-Magnus expansion is related to Dyson series \cite{Dyson1949DysonSeries} and the (general) time-dependent perturbation theory \cite{Blanes2009MagnusExpansion}.
In this appendix, we emphasize this point, by showing the similarity between observables calculated in the context of time-dependent perturbation theory and with the effective Hamiltonian in the context of the HFE, by considering the particular case of the anisotropic Haldane model.

\subsection{Perturbation theory}

We consider the system evolving with the Hamiltonian 
$\hat{H}(t) = \hat{H}_0 + \hat{V}(t)$ and we treat the time-dependent term perturbatively. 
We now define the operator
\begin{equation}
\hat{S}(t, t_0) = e^{i H_0 \left(t-t_0\right)}\hat{U}(t,t_0)
\end{equation}
where
\begin{equation}
\hat{U}(t,t_0) = 
\mathcal{T} \exp \bigg[ -i \int^t_{t_0} dt_1 \hat{H}(t_1) \bigg]
\end{equation}
is the evolution operator and $\mathcal{T}$ the time-ordered product.
One easily obtains that
\begin{equation}
\label{eq:app:S}
\hat{S}(t,t_0) = 
\mathcal{T} \exp \bigg[-i \int^t_{t_0} dt_1 \hat{V}_I(t_1, t_0) \bigg]
\end{equation}
where $\hat{A}_I(t, t_0)$ is the operator in the interaction picture:
$ \hat{A}_I(t, t_0) = e^{i \hat{H}_0 \left(t-t_0\right)}
\hat{A}(t) e^{-i \hat{H}_0 \left(t-t_0\right)} $.
We consider that the argument in the exponential in Eq.~\eqref{eq:app:S} is small.
We thus expand the exponential in Dyson series and calculate the time-ordered product:
\begin{align}
\hat{S}(t,t_0) = 
1 &- i \int^t_{t_0} \hat{V}_I(t_1,t_0) dt_1 
\notag \\  & -
\int^t_{t_0} dt_1 \int^{t_1}_{t_0} dt_2 
	\hat{V}_I(t_1,t_0) \hat{V}_I(t_2,t_0) + \dots
\end{align}
When the system is initially prepared in state $\ket{\phi(t=0)} = \ket{\phi_0}$, the mean value of an observable $O$ at time $t$ reads
\begin{equation}
O(t) = \braket{\hat{O}(t)}_t = 
\bra{\phi_0} \hat{S}^\dag(t,t_0) \hat{O}_I(t,t_0) \hat{S}(t,t_0) \ket{\phi_0}\;.
\end{equation}
This provides an expansion of $O(t)$ and study the first contributions in the following.

\subsection{Observables}

\subsubsection{Second neighbor currents}

The continuity equation in the Schr\"odinger picture for the particle density $n_{A,i}$ at site $i$ on sublattice $A$ takes the form
\begin{equation}
\frac{d \braket{\hat{n}_{A,i}}_t}{d t} =
\frac{d}{dt} \bra{\phi(t)} \hat{n}_{A,i} \ket{\phi(t)}  =
-\braket{\hat{J}_{A,i}(t)}_t
\end{equation}
The operator $\hat{J}_{A,i}(t)$ represents the sum of local instantaneous currents at the corresponding site. When the system is out-of-equilibrium, these currents do not vanish in general. By performing the calculation of the commutator, we immediately obtain:
\begin{equation}
\hat{J}_{A,i}(t) = i\sum_{j(i)}
\left[ -t_1 + V\cos\left( \omega t + \theta_{ij} \right) \right] 
\left( \hat{a}^\dag_i \hat{b}_j - \hat{b}^\dag_j \hat{a}_i \right)
\end{equation}
where $\sum_{j(i)}$ denotes the sum over all sites $j$, NN of a given site $i$.

After some calculations, one gets that the first non-trivial term, that does not vanish when integrated over an integer number of periods, appears at first order in the Dyson expansion as follows:
\begin{equation}
{J}_{A,i}^{(2)}(t) = 
\Braket{
i \bigg[ \int^t_{t_0} dt_1 \hat{V}_I(t_1,t), 
\hat{J}_{A,i}^{(1)}(t) \bigg] 
}_{t,0}
\end{equation}
where $\Braket{\cdots}_{t,0} = \Bra{\phi_0} e^{i\hat{H}_0\left(t-t_0\right)} 
\left(\cdots\right) e^{-i\hat{H}_0\left(t-t_0\right)} \Ket{\phi_0}$.
Now we want to take into account the fact that $\omega$ is the largest energy scale in the system. 
We perform the integration over time and neglect all subdominant order terms in $1/\omega^2$.
We obtain
\begin{widetext}
\begin{equation}
\label{eq:crntsIntPicture}
J_{A,i}^{(2)}(t) \approx
-\frac{V^2}{\omega} \bigg[
\sum_{j(i), l(i)}
\cos\left(\omega t + \theta_{ij} \right)
\sin\left(\omega t + \theta_{il} \right)
\braket{\hat{b}^\dag_j \hat{b}_l + \hat{b}^\dag_l \hat{b}_j}_{t,0}
-\sum_{\braket{\braket{k|i}}}
\cos\left(\omega t + \theta_{ij} \right)
\sin\left(\omega t + \theta_{kj} \right)
\braket{\hat{a}^\dag_i \hat{a}_k + \hat{a}^\dag_k \hat{a}_i}_{t,0}
\bigg]
\end{equation}
We notice that the first term is odd under the change $j \rightarrow l$ and that $j$ is fixed by the choice of $i$ and $k$ in the second term.
We are interested in calculating the average of $J_{A,i}^{(2)}(t)$ over the period $T$. After some calculations this leads to:
\begin{equation}
\int\limits_{t_0}^{t_0+T} dt \frac{d \braket{\hat{n}_{A,i}}_t}{d t} = 
\frac{V^2}{2\omega}
\sum\limits_{\braket{\braket{k|i}}}
\sin\left( \Theta_{ik} \right)
\int\limits_{t_0}^{t_0+T} dt
\braket{\hat{a}^\dag_i \hat{a}_k + \hat{a}^\dag_k \hat{a}_i}_{t,0}
\end{equation}
\end{widetext}
Here $\Theta_{ik} = \theta_{ij} + \theta_{jk}$ for any NNN sites $i$ and $k$ on the sublattice $A$ with the same NN site $j$ on the sublattice $B$. 
The first term has droped in the expression because of the sign under the exchange $j \rightarrow l$.
Last, we notice that we recover the expression of the $t_{2,ik}$ hopping amplitude found in the HFE.

Finally, we see that these averaged currents are almost the same as the ones obtained from the effective Hamiltonian. The only difference is that effective currents are calculated in the state evolving with the complete effective Hamiltonian $\hat{H}_\text{eff}$, in which case we take the average $\braket{\cdots}_t$, while averaged currents are calculated in the state evolving with $\hat{H}_0$, in which case the average is $\braket{\cdots}_{t,0}$:
\begin{align}
\frac{d \braket{\hat{n}^\text{eff}_{A,i}}_t}{d t} &=
-i \braket{ \big[ \hat{n}_{A,i}, \hat{H}_\text{eff}^{(1)} \big] }_t 
\notag \\ &=
\frac{V^2}{2 \omega}
\sum\limits_{\braket{\braket{k|i}}}
	\sin \left( \Theta_{ik} \right)
	\braket{\hat{a}^\dag_i \hat{a}_k + \hat{a}^\dag_k \hat{a}_i}_t
\end{align}

\subsubsection{Displacement of the momentum distribution}

The variation of the momentum distribution ${n}_{\mathbf{k}}$ is closely (but not directly) related to the form of currents expressed in the previous subsection. 
In the initial state, the momentum distribution is likely to be peaked at $\mathbf{k}=0$ so observing a FM condensate requires that ${n}_{\mathbf{k}}$ evolves significantly in time. 
More precisely, we have on sublattice $A$ the contribution
\begin{equation}
\frac{d \braket{\hat{n}_{A, \mathbf{k}}}_t}{dt} = 
\frac{1}{N_c} \sum\limits_i 
\bigg( 
	\frac{d \braket{\hat{n}_{A, i}}_t}{dt} +
	\sum\limits_{j\neq i} e^{i\mathbf{k}\left(\mathbf{r}_i-\mathbf{r}_j\right)}
	\frac{d \braket{\hat{a}^\dag_{i} \hat{a}_{j}}_t}{dt}
\bigg)
\end{equation}
Thanks to the fact that effective NNN currents should be conserved, one can argue that the first term in the expression above should sum up to zero when integrated over the full period of time $T$.

In order to calculate the effect of the second term, we need to perform the power expansion of 
$d \braket{\hat{a}^\dag_{i} \hat{a}_{j}}_t / dt$ at the same order as $d \Braket{\hat{n}_{A, i}}_t / dt$. We see that despite the more complicated dependency on the on-site creation and annihilation operators $\hat{a}^\dag_i(\hat{b}^\dag_i)$ and $\hat{a}_i(\hat{b}_i)$ (containing terms of currents between second and fourths nearest neighbours), this term will have exactly the same time-dependency through the factor 
\begin{equation}
\frac{V^2}{\omega}\sum\limits_{\braket{\braket{p|j}}}
\left[ \cos\left(\omega t + \theta_{jk} \right)
\sin\left(\omega t + \theta_{pk} \right) \right]
\braket{\hat{a}^\dag_{p} \hat{a}_{i} + \hat{a}^\dag_{i} \hat{a}_{p}}_{t,0}
\end{equation}
When integrated over the full period of time, this factor will become an odd function of $\Theta_{jp} = \theta_{jk} + \theta_{kp}$, that sum up to zero inside the summation $\sum_{i<j}$.
Thus, we conclude that in the regime of weak perturbation, we should have 
\begin{equation}
\int\limits_{t_0}^{t_0+T} dt \frac{d \braket{\hat{n}_{A,\mathbf{k}}}_t}{d t} = 0
\end{equation}
up to the order $\left(V/\omega\right)^2$, irrelevant in the regime $\omega \rightarrow \infty$.

\subsubsection{Momentum distribution}

In the same way one can use Dyson series to calculate the expectation value of the momentum distribution $\hat{n}_{A, \mathbf{k}}$. In this case, the first nontrivial terms, that does not vanish after taking the integration over an integer number of periods, appears only in the second order in the expansion. It is written as follows:
\begin{align}
\label{eq:exValObs_2ndOrder}
n^{(2)}_{A, \mathbf{k}}(t) = -&
\Braket{ 
	\int^t_{t_0} dt_1 \int^{t_1}_{t_0} dt_2 
	\left\lbrace\hat{V}_I(t_1,t) \hat{V}_I(t_2,t), 
	\hat{n}_{A, \mathbf{k}}\right\rbrace
 }_{t,0}
\notag \\ +&
\Braket{
\left[ \int^t_{t_0} dt_1 \hat{V}_I(t_1,t) \right]
\hat{n}_{A, \mathbf{k}}
\left[ \int^t_{t_0} dt_1 \hat{V}_I(t_1,t) \right]
}_{t,0}
\end{align}
with $\left\lbrace,\right\rbrace$ the anticommutator.
One easily sees that different terms in the expression contribute as $\left(V/\omega\right)^2$, that is $1/\omega$ times smaller then the biggest contribution for currents:
\begin{equation}
\begin{split}
\Braket{\hat{n}_{A, \mathbf{k}}}_t =
\Braket{\hat{n}_{A, \mathbf{k}}}_{t,0} +
\frac{1}{\omega}\frac{V^2}{\omega} (\cdots)
\end{split}
\end{equation}\\
We conclude that the momentum distribution is not the most convenient observable to be studied in the weak perturbation regime since it does not show any effect. Yet, the time-dependent perturbation theory is not suitable to study the regime when the perturbation become strong, which is the usual one in our Floquet scheme.

\begin{widetext}
\section{FM phase in the intermediate regime \label{app:fmPhaseInter}}

We recall the approximation based on the consideration of the two-well structure, that we made in order to rewrite the Hamiltonian \eqref{eq:boseHubHam} in the intermediate regime of the \textit{FM} phase:
\begin{equation}
\begin{split}
\hat{a}_i &\approx \frac{
	e^{-i \bm{K}_c \bm{r}_i} \hat{a}_{ \bm{K}_c} +
	e^{ i \bm{K}_c \bm{r}_i} \hat{a}_{-\bm{K}_c}
	}{\sqrt{N_c}} \\ 
\hat{b}_i &\approx \frac{
	e^{-i \bm{K}_c \bm{r}_i} \hat{b}_{ \bm{K}_c} +
	e^{ i \bm{K}_c \bm{r}_i} \hat{b}_{-\bm{K}_c}
	}{\sqrt{N_c}}
\end{split}
\end{equation}
By using this, we obtain the most general approximation of the Hamiltonian:
\begin{equation}
\begin{split}
\hat{H} \approx &
\epsilon_0 N - 4 t_2 \sin\left( z_c \frac{\pi}{2} \right)
\sum\limits_{\mu = \pm}
\mu \left( 
\hat{a}^\dag_{ \mu \bm{K}_c} \hat{a}_{ \mu \bm{K}_c} -
\hat{b}^\dag_{ \mu \bm{K}_c} \hat{b}_{ \mu \bm{K}_c}
\right) - t_1
\left[ 2 \cos\left(\frac{\pi z_c}{2}\right) + 1 \right]
\sum\limits_{\mu = \pm} 
\left(
\hat{a}^\dag_{ \mu \bm{K}_c} \hat{b}_{ \mu \bm{K}_c} +
\text{h.c.} \right) \\ + &
	\frac{U}{2} \sum\limits_{\mu = \pm} \left[
\hat{a}^\dag_{\mu \bm{K}_c} \hat{a}_{\mu \bm{K}_c}
\left( 
\frac{\hat{a}^\dag_{\bm{K}_c} \hat{a}_{\bm{K}_c} + 
\hat{a}^\dag_{-\bm{K}_c} \hat{a}_{-\bm{K}_c}}{N_c} - 1
\right) +
\hat{b}^\dag_{\mu \bm{K}_c} \hat{b}_{\mu \bm{K}_c}
\left( 
\frac{\hat{b}^\dag_{\bm{K}_c} \hat{b}_{\bm{K}_c} + 
\hat{b}^\dag_{-\bm{K}_c} \hat{b}_{-\bm{K}_c}}{N_c} - 1
\right) \right] \\ + &
	\frac{U}{N_s} \left[
\left( \hat{a}^\dag_{ \bm{K}_c} \hat{a}_{-\bm{K}_c} \right)
\left( \hat{a}^\dag_{-\bm{K}_c} \hat{a}_{ \bm{K}_c} \right) +
\left( \hat{b}^\dag_{ \bm{K}_c} \hat{b}_{-\bm{K}_c} \right)
\left( \hat{b}^\dag_{-\bm{K}_c} \hat{b}_{ \bm{K}_c} \right) +
\text{h.c.}
\right]
\end{split}
\end{equation}
For the incoming purpose we define
\begin{equation}
\braket{\hat{a}_{ \pm \bm{K}_c}} =
\sqrt{N_{A,\pm}}e^{i\theta_{A,\pm}}
, \qquad
\braket{\hat{b}_{ \pm \bm{K}_c}} =
\sqrt{N_{B,\pm}}e^{i\theta_{B,\pm}}
\end{equation}
such that $N_{A,+} + N_{A,-} = N_A $, $N_{B,+} + N_{B,-} = N_B$,
$N_{A,\pm} + N_{B,\pm} = N_\pm $ and $N_A + N_B = N_+ + N_- = N$.
We notice that in the regime of two decoupled sublattices $t_2 \gg t_1$ one can simply write $N_{A,-} = N_{B,+} = 0$. Finally, we get the expression of the GS energy $E_\text{GS}$ and currents $J_{AA, \bm{v}_j}$ in terms of these new quantities:
\begin{align}
\label{eq:eGsCsf}
E_{GS} =& \epsilon_0 N -
	4 t_2 \sin\left( z_c \frac{\pi}{2} \right) \sum\limits_{\mu = \pm}
\mu \left( N_{A,\mu} - N_{B,\mu} \right) -
	2 t_1 \left[ 
2 \cos\left(\frac{\pi z_c}{2}\right) + 1 \right]
\sum\limits_{\mu = \pm}
\sqrt{N_{A,\mu}N_{B,\mu}} \cos \left(\theta_{B,\mu}-\theta_{A,\mu}\right)
\notag \\ +&
	\frac{U}{2} \left[
N_A \left( \frac{N_A}{N_c} - 1 \right) +
N_B \left( \frac{N_B}{N_c} - 1 \right)
\right] +
	\frac{U}{N_c}
\left( N_{A,+} N_{A,-} + N_{B,+} N_{B,-} \right)
\end{align}
\begin{align}
J_{AA, \bm{v}_1} = J_{AA, \bm{v}_2} &= -t_2\frac{N_A}{N_c}
 \cos \left( z_c \frac{\pi}{2} \right)
\notag \\
J_{AA, \bm{v}_3} &= 0
\end{align}
We see that the $t_1$-term pins to zero the difference between phases 
$\theta_{B,+}-\theta_{A,+}$ and 
$\theta_{B,-}-\theta_{A,-}$. 
However, we are still free to chose separately phases 
$\theta_+ = \theta_{A,+} = \theta_{B,+}$ and 
$\theta_- = \theta_{A,-} = \theta_{B,-}$, which corresponds again to the presence of two Goldstone modes.

We assume that the effect of interactions is weak enough so that we use the assumptions of Eq.~\eqref{eq:csfGs}, ie. that the two-well structure of the system is preserved. We however allow for the fact that interactions can change the value of $z_c$ and modify the position of two minima at $\pm\bm{K}_c$. In order to study in more details this effect, we use the relations between operators $\hat{a}^\dag_{ \bm{K}_c}$ and $\hat{a}^\dag_{-\bm{K}_c}$, based on properties of the unitary transformation \eqref{eq:unitaryTransformationZc}, and write the GS energy in terms of $N$, $N_+$, $N_-=N-N_+$ and $z_c$ only:
\begin{align}
\label{eq:eGsCsfIntEffects}
E_{GS} =& 
N \left\lbrace
	\epsilon_0  -
	4 t_2 \sin\left(z_c \frac{\pi}{2}\right) \frac{Y(z_c)}{\sqrt{X^2(z_c)+Y^2(z_c)}} -
	t_1 \left[2 \cos\left(z_c \frac{\pi}{2}\right) + 1\right]
	\frac{X(z_c)}{\sqrt{X^2(z_c)+Y^2(z_c)}}
\right\rbrace
\notag \\ +&
	U N \left[\frac{n}{2} + \frac{Y^2(z_c)}{X^2(z_c)+Y^2(z_c)}\frac{n}{2} - \frac{1}{2} \right] +
	U \left[\frac{N_+\left(N-N_+\right)}{N_s}\right]
	\left[\frac{X^2(z_c)-2Y^2(z_c)}{X^2(z_c)+Y^2(z_c)}\right]
\end{align}
where $X(z_c)$ and $Y(z_c)$ are defined as follows:
\begin{align}
X(z_c) &= t_1 \left[ 1 + 2 \cos\left( z_c \frac{\pi}{2} \right) \right]
\notag \\
Y(z_c) &= 4 t_2 \sin\left( z_c \frac{\pi}{2} \right)
\end{align}
The first effect of interactions in the \textit{FM} consists in imposing constraints on the distribution of particles in two wells $N_+$ and $N_-$. There are two distinct sub-regimes of the \textit{FM} phase: for $t_2^c < t_2 <  \sqrt{(17+\sqrt{97})/24}\ t_2^c$ the last term in Eq.~\eqref{eq:eGsCsfIntEffects} is positive and interactions force all particles to chose one particular well, such that $N_+$ or $N_-$ becomes precisely equal to $N$. At the other side, for $t_2 >  \sqrt{(17+\sqrt{97})/24}\ t_2^c$ the last term in Eq.~\eqref{eq:eGsCsfIntEffects} is negative and the uniform distribution of particles $N_+ = N_- = N/2$ is preferred.
The second effect of interactions consists in moving the position of minima at $\pm \bm{K}_c$.
The contribution of interactions to the GS energy close to the transition is
\begin{align}
\Delta E'_{GS} = 
NUn\frac{8 t_2^2}{9t_1^2} \left(z_c\frac{\pi}{2}\right)^2 +
N \frac{Un}{2} \left[
\frac{16}{27} \left(\frac{t_2}{t_1}\right)^2 -
\frac{256}{81} \left(\frac{t_2}{t_1}\right)^4
\right] \left(z_c\frac{\pi}{2}\right)^4
\end{align} 
If particles are not all located in one well, there appears an additional contribution
\begin{align}
\Delta E''_{GS} = -
\frac{16U}{3}
\frac{N_+\left(N-N_+\right)}{N_s}
\left(\frac{t_2}{t_1}\right)^2
\left(z_c\frac{\pi}{2}\right)^2 -
\frac{16U}{27}
\frac{N_+\left(N-N_+\right)}{N_s}
\left( \frac{3t_1^2t_2^2-16t_2^4}{t_1^4} \right)
\left(z_c\frac{\pi}{2}\right)^4
\end{align}
The second contribution $\Delta E''_{GS}$ dominates when $N_+ = N_- = N/2$, resulting in the generation of an effective repulsion between two wells.
\end{widetext}


\pagebreak
\section*{References}

%

\end{document}